%% file: Springer_20100722.tex
\documentclass[graybox]{svmult}

\usepackage{mathptmx}       % selects Times Roman as basic font
\usepackage{helvet}         % selects Helvetica as sans-serif font
\usepackage{courier}        % selects Courier as typewriter font
\usepackage{type1cm}        % activate if the above 3 fonts are
\usepackage{makeidx}         % allows index generation
\usepackage{graphicx}        % standard LaTeX graphics tool
\usepackage{multicol}        % used for the two-column index
\usepackage[bottom]{footmisc}% places footnotes at page bottom

\usepackage{xspace,units}
\usepackage{float}
\usepackage{amsmath}
\usepackage{amssymb}
\usepackage[normalem]{ulem}
\usepackage{wrapfig}
\usepackage{epsfig}
\usepackage{bm}
\definecolor{MidnightBlue}{cmyk}{0.98,0.13,0,0.43}%%%PANTONE 302
 \definecolor{DarkGreen}{rgb}{0,0.7,0.1}

\newcommand{\refeq}[1]{{(\ref{eq:#1})}}
\newcommand{\refeqn}[1]{{Eq.~(\ref{eq:#1})}}
\newcommand{\labeleqn}[1]{\label{eq:#1}}
\newcommand{\reffig}[1]{{Fig.~\ref{fig:#1}}}
\newcommand{\be}{\begin{equation}}
\newcommand{\ee}{\end{equation}}

\newcommand{\Lag}{\mathcal{L}}

\newcommand{\curl}{\boldsymbol{\nabla} \times}
\newcommand{\bnabla}{\boldsymbol{\nabla}}
\newcommand{\bfE}{\mathbf{E}}
\newcommand{\bfD}{\mathbf{D}}
\newcommand{\B}{\mathbf{B}}
\newcommand{\Hf}{\mathbf{H}}
\newcommand{\A}{\mathbf{A}}
\newcommand{\J}{\mathbf{J}}

\newcommand{\tV}{\mathbb{V}}
\newcommand{\bI}{\mathcal{I}}
\newcommand{\tI}{\mathbb{I}}
\newcommand{\tA}{\mathbb{A}}
\newcommand{\tB}{\mathbb{B}}
\newcommand{\tC}{\mathbb{C}}
\newcommand{\tD}{\mathbb{D}}
\newcommand{\calE}{\mathcal{E}}

\newcommand{\minmunew}{\left(\frac{1}{\mu(\omega,\vecx)}-1\right)}

\newcommand{\minepnew}{\left(1-\epsilon(\omega,\vecx)\right)}
\newcommand{\ep}{\epsilon}

\newcommand{\dA}{\mathcal{D}\A}

\newcommand{\dJ}{\mathcal{D}\J}
\newcommand{\dJJ}{\left. \dJ\dJ^* \right|_\text{obj}}
\newcommand{\dJJprime}{\left. \dJ'{\dJ'}^{*} \right|_\text{obj}}
\newcommand{\bra}[1]{\langle #1 |}
\newcommand{\ket}[1]{| #1 \rangle}

\newcommand{\tr}{\text{tr }}
\newcommand{\half}{\frac{1}{2}}

\newcommand{\aindex}{\alpha}
\newcommand{\bindex}{\beta}
\newcommand{\cindex}{\gamma}
\newcommand{\abindex}{\aindex \bindex}

\newcommand{\F}{\mathbb{F}}
\newcommand{\T}{\mathbb{T}}
\newcommand{\f}{\mathcal{F}}

\newcommand{\X}{\mathbb{X}}

\newcommand{\V}{\mathbb{V}}
\newcommand{\Y}{\mathcal{N}}
\newcommand{\Hzero}{\mathbb{H}_0}
\newcommand{\Ham}{H}

\newcommand{\bfEh}{\bfE_0}
\newcommand{\bfEin}{\bfE^\text{in}}
\newcommand{\bfEout}{\bfE^\text{out}}
\newcommand{\bfEreg}{\bfE^\text{reg}}

\newcommand{\bfEregcc}{\bfE^{\text{reg}*}}
\newcommand{\bfEregbQcc}{\bfEregcc_{\bindex}}
\newcommand{\bfEregaP}{\bfEreg_{\aindex}}
\newcommand{\bfEoutaP}{\bfEout_{\aindex}}
\newcommand{\bfEoutkaaP}{\bfEout_{\aindex}(\kappa)}
\newcommand{\bfEoutkabQ}{\bfEout_{\bindex}(\kappa)}
\newcommand{\bfEinaPcc}{\bfE_{\aindex}^{\text{in}*}}
\newcommand{\bfEinbQcc}{\bfE_{\bindex}^{\text{in}*}}

\newcommand{\bfEinkaaP}{\bfEin_{\aindex}(\kappa)}
\newcommand{\bfEinkabQ}{\bfEin_{\bindex}(\kappa)}

\newcommand{\bfEoutbQ}{\bfEout_{\bindex}}

\newcommand{\bfEom}{\bfE(\omega)}

\newcommand{\bfEhom}{\bfEh(\omega)}
\newcommand{\bfEromaP}{\bfE_{\aindex}^{\text{reg}}(\omega)}

\newcommand{\bfEoutomaP}{\bfEout_{\aindex}(\omega)}
\newcommand{\bfEoutombQ}{\bfEout_{\bindex}(\omega)}
\newcommand{\bfErombQ}{\bfEr_{\bindex}(\omega)}
\newcommand{\bfEinombQ}{\bfEin_{\bindex}(\omega)}

\newcommand{\ketEom}{\ket{\bfEom}}

\newcommand{\ketEromaP}{\ket{\bfEromaP}}

\newcommand{\ketEoutomaP}{\ket{\bfEoutomaP}}
\newcommand{\ketEoutombQ}{\ket{\bfEoutombQ}}
\newcommand{\ketErombQ}{\ket{\bfErombQ}}
\newcommand{\braErombQ}{\bra{\bfErombQ}}
\newcommand{\braEinombQ}{\bra{\bfEinombQ}}
\newcommand{\braEinkaaP}{\bra{\bfEinkaaP}}

\newcommand{\ketE}{\ket{\bfE}}
\newcommand{\ketEh}{\ket{\bfEh}}

\newcommand{\ketEoutkaaP}{\ket{\bfEoutkaaP}}

\newcommand{\braEinkabQ}{\bra{\bfEinkabQ}}

\newcommand{\ketEoutkabQ}{\ket{\bfEoutkabQ}}

\newcommand{\tG}{\mathbb{G}}
\newcommand{\tGzero}{\tG_0}
\newcommand{\tGM}{\mathbb{G}_M}
\newcommand{\bfEr}{\bfE^\text{reg}}

\newcommand{\bfEraP}{\bfE_{\aindex}^{\text{reg}}}
\newcommand{\bfErkaaP}{\bfE_{\aindex}^{\text{reg}}(\kappa)}
\newcommand{\bfEraPcc}{\bfE_{\aindex}^{\text{reg}*}}

\newcommand{\bfErkabQ}{\bfEr_{\bindex}(\kappa)}

\newcommand{\ketErkaaP}{\ket{\bfErkaaP}}

\newcommand{\ketErkabQ}{\ket{\bfErkabQ}}

\newcommand{\braErkabQ}{\bra{\bfErkabQ}}

\newcommand{\braErkaaP}{\bra{\bfE_{\aindex}^{\text{reg}}(\kappa)}}

\newcommand{\veck}{\mathbf{k}}
\newcommand{\veckpe}{\veck_\perp}
\newcommand{\vecd}{\mathbf{d}}

\newcommand{\vecx}{\mathbf{x}}

\newcommand{\vecX}{\mathbf{X}}

\newcommand{\orig}{\mathcal{O}}

\newcommand{\ketxp}{|\vecx'\rangle}
\newcommand{\ketxpp}{|\vecx''\rangle}

\newcommand{\ketxip}{\ket{\vecx_i'}}
\newcommand{\ketxjpp}{\ket{\vecx_j''}}
\newcommand{\ketxonep}{|\vecx_1'\rangle}
\newcommand{\ketxtwop}{|\vecx_2'\rangle}
\newcommand{\ketxonepp}{|\vecx_1''\rangle}
\newcommand{\ketxtwopp}{|\vecx_2''\rangle}
\newcommand{\brax}{\langle \vecx|}
\newcommand{\braxi}{\bra{\vecx_i}}
\newcommand{\braxip}{\bra{\vecx_i'}}
\newcommand{\braxone}{\langle \vecx_1|}
\newcommand{\braxtwo}{\langle \vecx_2|}
\newcommand{\braxonep}{\langle \vecx_1'|}
\newcommand{\braxtwop}{\langle \vecx_2'|}

\newcommand{\Tregreg}{\f^{ee}}

\newcommand{\Tregout}{\f^{ei}}
\newcommand{\Toutreg}{\f^{ie}}
\newcommand{\Toutout}{\f^{ii}}

\newcommand{\ka}{\kappa}

\newcommand{\natexlabmod}[1]{}

\def\aadd{s}
 \def\badd{t}
\def\adel{s}
 \def\bdel{t}
\def\acomm{s}
 \def\bcomm{t}

 \newcommand{\add}[1]{\if\aadd\badd{{\color{BrickRed}#1}}\else{#1}\fi}
\newcommand{\del}[1]{\if\adel\bdel{{\color{Gray}[[#1]]}}\else{}\fi}
\newcommand{\comm}[1]{\if\acomm\bcomm{{\color{MidnightBlue}\{#1\}}}\else{}\fi}
\makeindex             % used for the subject index
\begin{document}

\title*{Geometry and material effects in Casimir physics - Scattering theory}
% Use \titlerunning{Short Title} for an abbreviated version of
% your contribution title if the original one is too long
\author{Sahand Jamal Rahi, Thorsten Emig, and  Robert L. Jaffe}
% Use \authorrunning{Short Title} for an abbreviated version of
% your contribution title if the original one is too long
\institute{S. J. Rahi \at Department of Physics, MIT, Cambridge, MA, \email{sjrahi@mit.edu}
\and T. Emig \at Institut f\"ur Theoretische Physik, Universit\"at zu K\"oln,
 Z\"ulpicher Strasse 77, 50937 K\"oln, Germany and
 Laboratoire de Physique Th\'eorique et Mod\`eles
 Statistiques, CNRS UMR 8626, Universit\'e Paris-Sud, 91405 Orsay,
 France,
\email{emig@lptms.u-psud.fr}
\and R. L. Jaffe \at Center for Theoretical Physics\,MIT, Cambridge, MA, \email{jaffe@mit.edu}}
%
% Use the package "url.sty" to avoid
% problems with special characters
% used in your e-mail or web address
%
\maketitle

\abstract{We give a comprehensive presentation of methods for calculating the Casimir force to arbitrary accuracy, for any number of objects, arbitrary shapes, susceptibility functions, and separations. The technique is applicable to objects immersed in media other than vacuum, to nonzero temperatures, and to spatial arrangements in which one object is enclosed in another.  Our method combines each object's classical  electromagnetic scattering amplitude with universal translation matrices, which convert between the bases used to calculate scattering for each object, but are otherwise independent of the details of the individual objects.  This approach, which combines methods of statistical physics and scattering theory, is well suited to analyze many diverse phenomena.  We illustrate its power and versatility  by a number of examples, which show how the interplay of geometry and material properties helps to understand and control Casimir forces. We also examine whether electrodynamic Casimir forces can lead to stable levitation. Neglecting permeabilities, we prove that any equilibrium position of objects subject to such forces is unstable if the permittivities of all objects are higher or lower than that of the enveloping medium; the former being the generic case for ordinary materials in vacuum.
}

%(20 - 30 pages)

\input{chap1}

\input{chap2}

\input{chap3}

\section{Applications}
\label{sec:Applications}

This section gives an overview on different geometries and shapes that
have been studied by the approach that we introduced in Section \ref{sec:energy-from-field}.
A selection of applications has been made to showcase generic situations and important effects that
had not been studied in detail before the development of the methods described here.  We shall mainly summarize analytical and numerical results for the Casimir interaction in the various systems.  For details on their derivation and additional implementations of the scattering approach we refer to the literature.

\subsection{Cylinders, wires, and plates}
\label{subsec:1D}

The {extent} to which EM field fluctuations are correlated depends on
the effective dimensionality of the space that can be explored by the
fluctuations. Therefore, Casimir interactions are expected to depend
strongly on the codimension of the interacting objects.  The focus of
this {subsection} is on the particular properties of systems with a
codimension of the critical value two. We consider these problems in
the context of interactions between cylinders and a cylinder and a
plate, both perfect reflectors and dielectric materials.
Cylindrical geometries are of recent experimental interest since they
are easier to hold parallel than plates and still generate a force that is extensive in
one direction.

\begin{figure}[ht]
\sidecaption[t]
\includegraphics[width=0.6\linewidth]{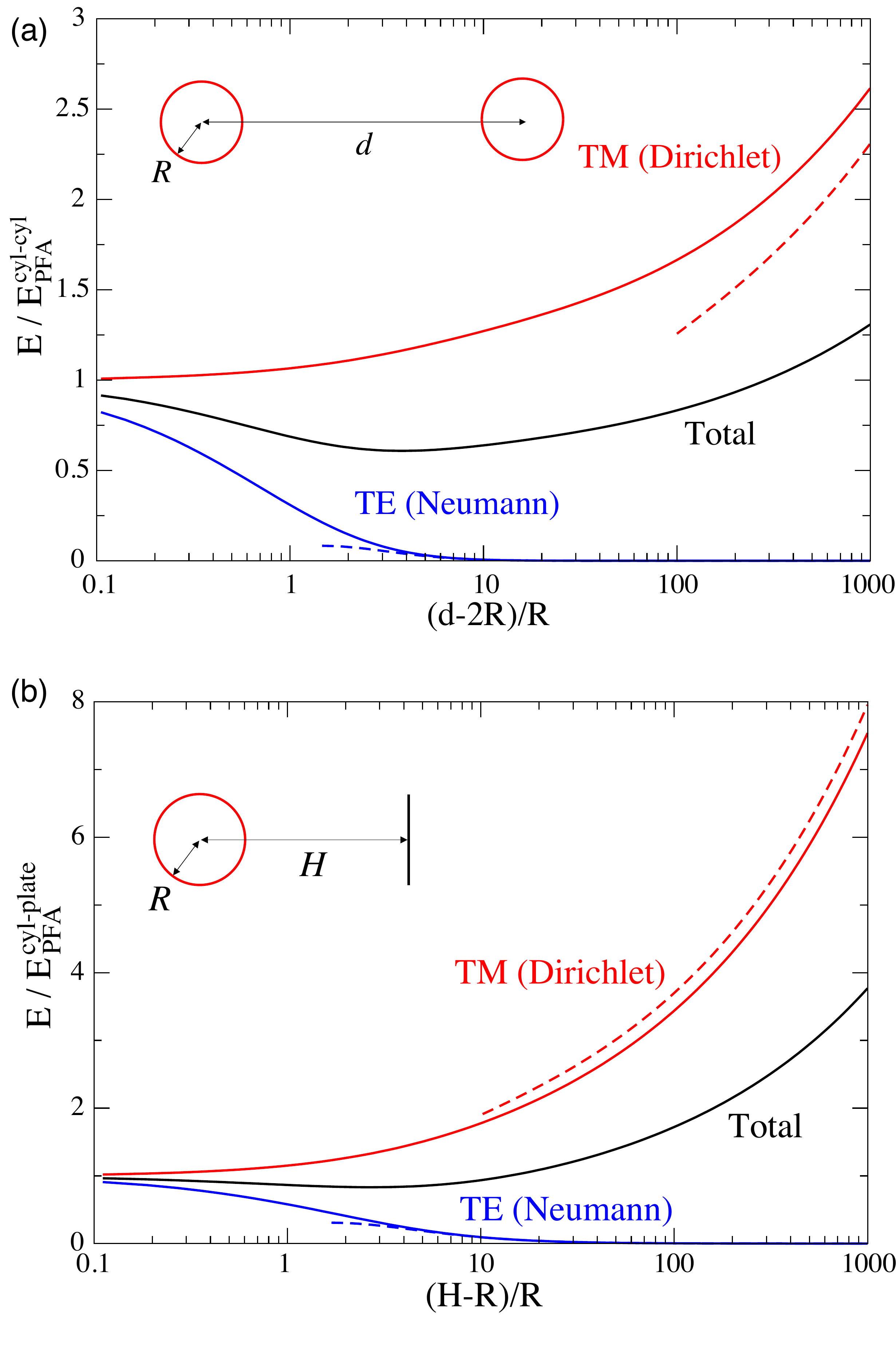}
\caption{(a) Casimir energy for two cylinders of equal radius $R$ as a
  function of surface-to-surface distance $d-2R$ (normalized by the
  radius). The energy is divided by the PFA estimate $E_{PFA}^{cyl-cyl}$ for
  the energy. The solid curves show our
  numerical results; the dashed lines represent the asymptotic results
  of Eq.~\eqref{eq:E-two-cylinders}. 
(b) Casimir energy for a cylinder of radius $R$ parallel to a
  plate as a function of the surface-to-surface distance $H-R$
  (normalized by the radius). The energy is divided by the PFA
  estimate $E_{PFA}^{cyl-plate}$. The
  solid curves reflect our numerical results; the dashed lines
  represent the asymptotic results of
  Eq. \eqref{eq:E-cyl-plate}. 
}
\label{fig:cyl_cyl}
\end{figure}

We consider two cylinders of equal radii $R$ and length $L\to\infty$
with center-to-center separation $d$, see
Fig.~\ref{fig:cyl_cyl}(a) \cite{Rahi08-2}.{ (The related configuration where one cylinder is inside another cylinder is treated in reference \cite{Dalvit06}.)}
For this geometry the interaction energy is
obtained from the expression
\begin{equation}
\begin{split} 
\label{eq:E_general}
  \mathcal{E} & = \frac{\hbar c}{2\pi} \int_0^\infty d\kappa \log\det
  \left(\bI-\Tregreg_\text{cyl} \mathcal{U}^{ba} \Tregreg_\text{cyl}
    \mathcal{U}^{ab} \right).
\end{split} 
\end{equation}
with the exterior scattering amplitudes of a cylinder,
\begin{equation}
\label{eq:F_cyl}
\begin{split} 
\Tregreg_{\rm{cyl},k_z' n' E,k_z n M} & = 
\Tregreg_{\rm{cyl},k_z' n' M,k_z n E} = 0, \\ 
\Tregreg_{\rm{cyl},k_z' n' M,k_z n M} & = 
-\tfrac{2\pi}{L}\delta(k_z-k_z')\delta_{n,n'} 
\frac{I'_n \left( R p\right)}{K'_n\left(R p\right)}\, , \\ 
\Tregreg_{\rm{cyl},k_z' n' E,k_z n E} & = 
-\tfrac{2\pi}{L}\delta(k_z-k_z')\delta_{n,n'} 
\frac{I_n \left( R p\right)}{K_n\left(R p\right)} \, ,
\end{split} 
\end{equation}
and the matrices $\mathcal{U}^{ab}$, $\mathcal{U}^{ba}$ that
translate from cylinder $a$ to $b$ and vice versa. Their elements are
summarized in Ref.~\cite{Rahi09}. The matrix inside the determinant is
diagonal in $k_z$, so the log-determinant over this index turns into
an overall integral.  A change of variable to polar coordinates
converts the integrals over $\kappa$ and $k_z$ to a single integral
over $p=\sqrt{k_z^2+\kappa^2}$, yielding \be \mathcal{E} = \frac{\hbar
  c L}{4\pi}\int_0^\infty pdp \left(\log\det \Y^M + \log\det
  \Y^E\right), \ee where \be
\begin{split} 
\Y^M_{n,n''} & = \delta_{n,n''} - \sum_{n'} 
\frac{I'_n(p R)}{K'_n(p R)} K_{n+n'}(pd) 
\frac{I'_{n'}(p R)}{K'_{n'}(p R)} K_{n'+n''}(pd) \\ 
\Y^E_{n,n''} & = \delta_{n,n''} - \sum_{n'} 
\frac{I_n(p R)}{K_n(p R)} K_{n+n'}(pd) 
\frac{I_{n'}(p R)}{K_{n'}(p R)} K_{n'+n''}(pd)
\end{split} 
\ee 
describe magnetic (TE) or Neumann modes and electric (TM) or
Dirichlet modes, respectively.

For large separations $d\gg R$, the asymptotic behavior of the energy
is determined by the matrix elements for $n=n'=0$ for Dirichlet modes and
$n=n'=0,\, \pm 1$ for Neumann modes. Taking the determinant of the matrix
that consists only of these matrix elements and integrating over $p$
yields straightforwardly the attractive interaction energies
\begin{equation}
\label{eq:E-two-cylinders}
\begin{split}
\mathcal{E}^E & = -\frac{\hbar c L}{d^2}\frac{1}{8\pi\log^2(d/R)} 
\left(1-\frac{2}{\log(d/R)} + \ldots \right) \, , \\
\mathcal{E}^M & = -\hbar c L \frac{7}{5\pi}\frac{R^4}{d^6} 
\end{split}
\end{equation}
for electric (Dirichlet) and magnetic (Neumann) modes.  The asymptotic
interaction is dominated by the contribution from electric (Dirichlet) modes
that vanishes for $R \to 0$ only logarithmically.

For arbitrary separations higher order partial waves have to be
considered. The number of partial waves has to be increased with
decreasing separation. A numerical evaluation of the determinant and
the $p$-integration can be performed easily and reveals an
exponentially fast convergence of the energy in the truncation order
for the partial waves.  Down to small surface-to-surface separations
of $(d-2R)/R=0.1$ we find that $n=40$ partial waves are sufficient to
obtain precise results for the energy. The corresponding result for
the energies of two cylinders of equal radius is shown in
Fig.~\ref{fig:cyl_cyl}(a). Notice that the minimum in the curve for
the total electromagnetic energy results from the scaling by the
proximity force approximation (PFA) estimate of the energy. The total
energy is monotonic and the force attractive at all separations.

Next we consider a cylinder and an infinite plate, both perfectly
reflecting, see Fig.~\ref{fig:cyl_cyl}(b). The Casimir energy for this
geometry has been computed originally in Ref.~\cite{Emig06}.  In the
limit of perfectly reflecting surfaces, the method of images can be
employed to compute the Casimir interaction for this geometry
\cite{Rahi08-2}. Here we use a different method that can be also
applied to real metals or general dielectrics \cite{Rahi09}. 
We express the scattering amplitude of the cylinder now
in a plane wave basis, using
\be 
\Tregreg_{\rm{cyl},\veckpe P,\veckpe' P'} 
 = \sum_{nQ,n'Q'} 
\frac{C_{\veckpe P}(\kappa)}{C_{Q}} 
D^\dagger_{\veckpe P, k_z n Q} 
\Tregreg_{\rm{cyl},k_z n Q,k_z n'Q'} 
D_{k_z n' Q', \veckpe' P'}, 
\labeleqn{Tcylinplate} 
\ee where $\veckpe$ denotes the vector $(k_y,k_z)$, $C_{\veckpe
  P}(\kappa)$ and $C_{Q}$ are normalization coefficients that can be
found together with the matrix elements of the conversion matrix $D$
in Ref.~\cite{Rahi09}. The elements of the scattering amplitude in the
cylindrical basis are given by Eq.~(\ref{eq:F_cyl}). The scattering
amplitude of the plate is easily expressed in the plane wave basis as
\begin{equation}
\begin{split} 
\Tregreg_{\rm{plate},\veckpe' E,\veckpe M} & = \Tregreg_{plate,\veckpe' M,\veckpe E} 
= 0 \, , \\ 
\Tregreg_{\rm{plate},\veckpe' M,\veckpe M} & = 
\tfrac{(2\pi)^2}{L^2}\delta^{(2)}(\veckpe-\veckpe') \, 
r^M\left(ic\kappa,\sqrt{1+\veckpe^2/\kappa^2}^{-1}\right) \, ,\\ 
\Tregreg_{\rm{plate},\veckpe' E,\veckpe E} & = 
\tfrac{(2\pi)^2}{L^2}\delta^{(2)}(\veckpe-\veckpe') \, 
r^E\left(ic\kappa,\sqrt{1+\veckpe^2/\kappa^2}^{-1}\right) \, , 
\end{split} 
\label{Tplate} 
\end{equation}
in terms of the Fresnel coefficients that read for a general
dielectric surface \be
\begin{split} 
r^M(ic\kappa,x) & = 
\frac 
{\mu(ic\kappa) - \sqrt{1+(n^2(ic\kappa)-1)x^2}} 
{\mu(ic\kappa) + \sqrt{1+(n^2(ic\kappa)-1)x^2}} \, , \\ 
r^E(ic\kappa,x) & = 
\frac 
{\epsilon(ic\kappa) - \sqrt{1+(n^2(ic\kappa)-1)x^2}} 
{\epsilon(ic\kappa) + \sqrt{1+(n^2(ic\kappa)-1)x^2}}. 
\end{split} 
\labeleqn{Fresnel} 
\ee Here, $n$ is the index of refraction,
$n(ic\kappa)=\sqrt{\epsilon(ic\kappa) \mu(ic\kappa)}$.  In the limit
of a perfectly reflecting plate one has $r^M\to -1$, $r^E\to 1$. The
energy given by Eq.~(\ref{eq:E_general}) can now be evaluated in the
plane wave basis with the translation matrices given by the simple
expression $\mathcal{U}^{ab}_{\veckpe P, \veckpe' P'} = e^{-
  \sqrt{\veckpe^2+\kappa^2}H}
\tfrac{(2\pi)^2}{L^2}\delta^{(2)}(\veckpe-\veckpe')\delta_{P,P'}$.

The asymptotic expression for the attractive interaction energy at
large distance $H\gg R$ reads
\begin{equation}
\begin{split}
\mathcal{E}^E & = -\frac{\hbar c L}{H^2}\frac{1}{16\pi\log(H/R)}, \\
\mathcal{E}^M & = - \hbar c L \frac{5}{32\pi} \frac{R^2}{H^4} \, .
\end{split}
\label{eq:E-cyl-plate}
\end{equation}
The total electromagnetic Casimir interaction is again dominated by
the contribution from the electric (Dirichlet) mode with $n=0$ which
depends only logarithmically on the cylinder radius.  The interaction
at all separations follows, as in the case of two cylinders, from a
numerical computation of the determinant of Eq.~(\ref{eq:E_general})
and integration over $p$.  The result is shown in
Fig.~\ref{fig:cyl_cyl}(b).

The above approach has the advantage that it can be also applied to
dielectric objects. The scattering amplitude of a dielectric cylinder
can be obtained by solving the wave equation in a cylindrical basis
with appropriate continuity conditions \cite{Rahi09}.  The scattering
amplitude is diagonal in $k_z$ and the cylindrical wave index $n$, but
not in the polarization. Here we focus on large distances $H \gg R$. 
Expanding the $\log\det$ in Eq.~(\ref{eq:E_general}), we obtain for
the interaction energy 
\be 
\mathcal{E} = -\frac{3 \hbar c L R^2}{128
  \pi H^4} \int_0^1 dx \frac{\epsilon_{\rm{cyl},0}-1}{\epsilon_{\rm{cyl},0}+1}
\left[(7 + \epsilon_{\rm{cyl},0} - 4 x^2)r^E(0,x)-(3+\epsilon_{\rm{cyl},0}) x^2
  r^M(0,x)\right], 
\ee 
if the zero-frequency magnetic permeability
$\mu_{cyl,0}$ of the cylinder is set to one. If we do not set $\mu_{\rm{cyl},0}$
equal to one, but instead take the perfect reflectivity limit for the
plate, we obtain 
\be 
\mathcal{E} = -\frac{\hbar c L R^2}{32 \pi H^4}
\frac{(\epsilon_{\rm{cyl},0}-\mu_{\rm{cyl},0})(
  9+\epsilon_{\rm{cyl},0}+\mu_{\rm{cyl},0}+\epsilon_{\rm{cyl},0}\mu_{\rm{cyl},0})}
{(1+\epsilon_{\rm{cyl},0})(1+\mu_{\rm{cyl},0})}.  
\ee
 
\begin{figure}[ht]
\sidecaption[t]
\includegraphics[width=0.6\linewidth]{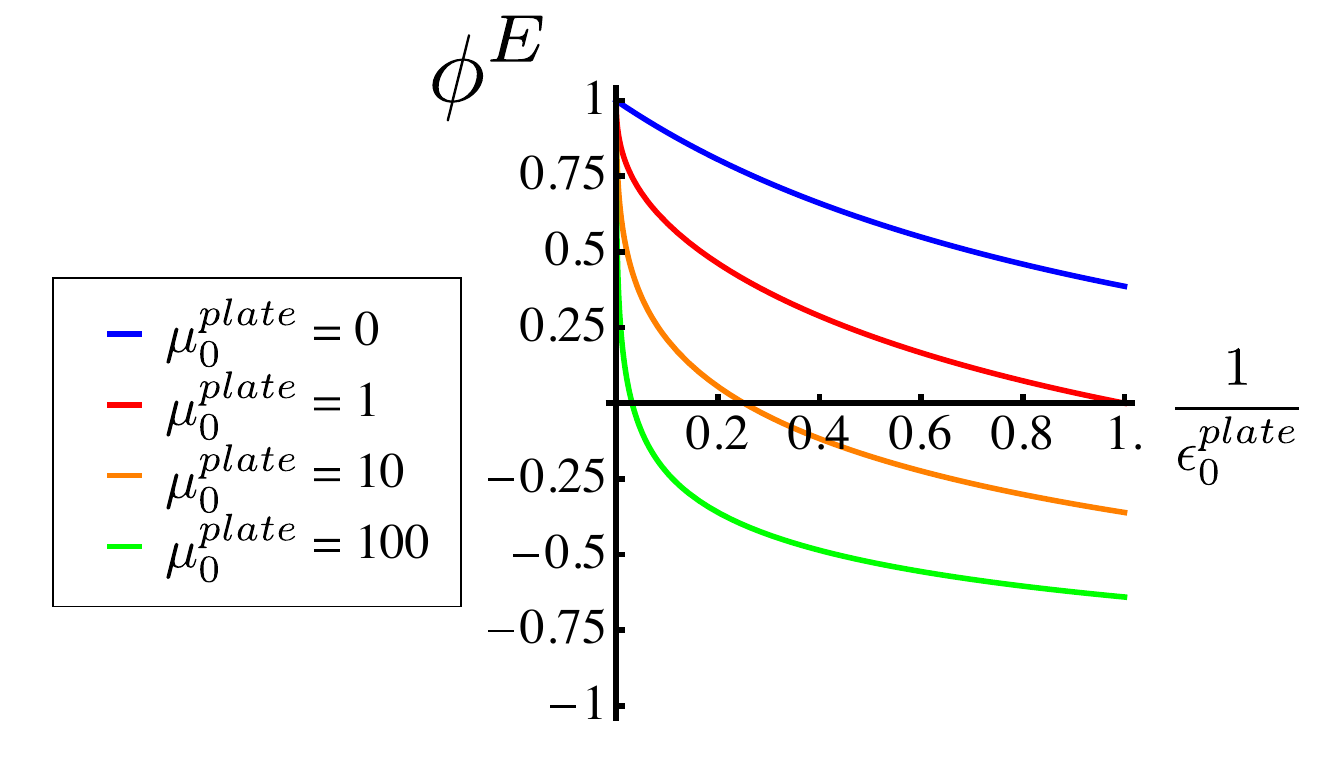}
\caption{
Plots of $\phi^E$ versus $1/\epsilon_{\rm{plate},0}$ for 
fixed values of $\mu_{\rm{plate},0}$. The perfect metal limit ($\phi^E=1$) is 
approached slowly for large $\mu_{\rm{plate},0}$, as in the case of a sphere 
opposite a plate. For large $\mu_{\rm{plate},0}$ the interaction becomes 
repulsive, which is expected given similar results for two infinite 
plates.
}
\label{fig:figplatecylasym}
\end{figure}

Finally, if we let $\epsilon_{\rm{cyl}}$ be infinite from the beginning
(the perfect metal limit for the cylinder), only the $n=0$ TM mode of
the scattering amplitude contributes at lowest order. For a plate with
zero-frequency permittivity $\epsilon_{\rm{plate},0}$ and permeability
$\mu_{\rm{plate},0}$, we obtain for the Casimir energy 
\be 
\mathcal{E} =
\frac{\hbar c L}{16 \pi H^2 \log(R/H)} \phi^E \, , \ee where \be
\phi^E = \int_0^1 \frac{dx}{1+x}\left[r^E(0,x)-x r^M(0,x)\right].
\ee 
In Fig.~\ref{fig:figplatecylasym}, $\phi^E$ is plotted as a function of the
zero-frequency permittivity of the plate, $\epsilon_{\rm{plate},0}$, for various
zero-frequency permeability values, $\mu_{\rm{plate},0}$.

\subsection{Three-body effects}
\label{subsec:3body}

\begin{figure}[ht]
%\sidecaption[t]
\begin{center}
\includegraphics[width=0.72\linewidth]{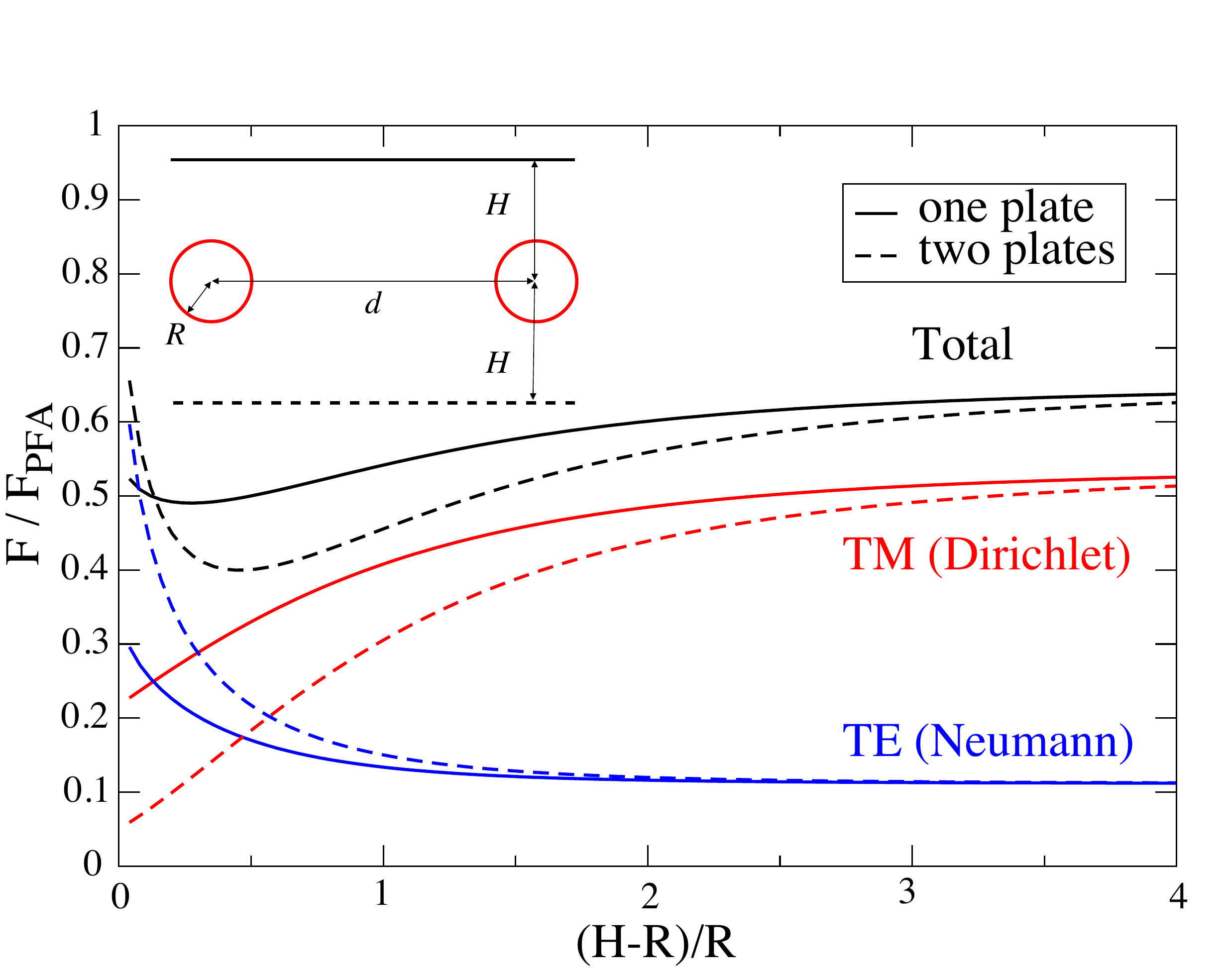}
\caption{ Electromagnetic Casimir force per unit length between two
  cylinders of radius $R$ and sidewall separation $H$ vs.  the ratio
  of sidewall separation to cylinder radius $(H-R)/R$, at fixed
  distance $(d-2R)/R=2$ between the cylinders, normalized by the total
  PFA force per unit length between two isolated cylinders,
  $F_{\text{PFA}}= \frac{5}{2}(\hbar c\pi^3/
  1920)\sqrt{R/(d-2R)^7}$. The force is attractive. The solid lines
  refer to the case with one sidewall, while dashed lines depict the
  results for two sidewalls.  Also shown are the individual TE (blue)
  and TM (red) forces. }
\label{fig:2cyl+plates}
\end{center}
\end{figure}

\begin{figure}[ht]
\begin{center}
%\sidecaption[t]
\includegraphics[width=0.9\linewidth]{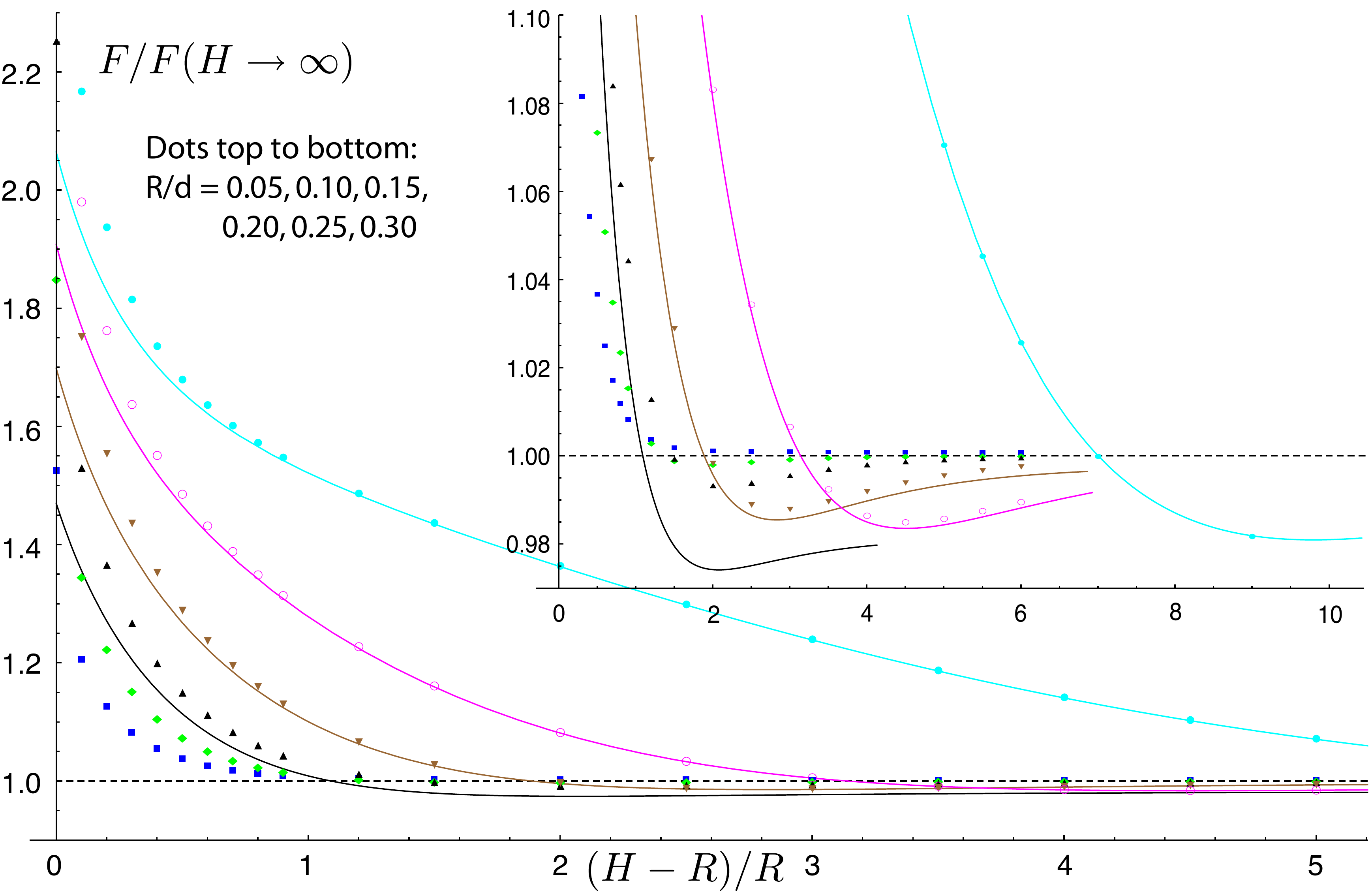}
\caption{ Electromagnetic Casimir force between two spheres next to
  one sidewall at separation $H$ vs. the ratio $H/R$ for different
  sphere separations $d$. Dotted curves represent numerical
  results. Shown are also the analytical results of
  Eq.~\eqref{eq:force-of-L}, including terms up to $j=10$ for $R/d\le
  0.2$ (solid curves)  \cite{Rodriguez09}. Inset: Magnification of the nonmonotonicity.  }
\label{fig:2spheres+plate}
\end{center}
\end{figure}

Casimir interactions are not pair-wise additive. To study the
consequences of this property, we consider the case of two identical
objects near perfectly reflecting walls~\cite{Rahi08-1,Rahi08-2}. Multibody effects were first observed for such a configuration with two rectangular cylinders sandwiched between two infinite plates by Rodriguez et al.~\cite{Rodriguez07:PRL}. The role of dimension on this
effect is studied by considering either cylinders, see
Fig.~\ref{fig:2cyl+plates}, or spheres, see
Fig.~\ref{fig:2spheres+plate}. While we have given a more detailed
description of how the interaction energies follow from the scattering
approach in the previous {subsection}, we mainly provide the final results
in this and in the following {subsections}.

First, we consider the geometry shown in Fig.~\ref{fig:2cyl+plates}
with two cylinders that are placed parallel to one or in-between two
parallel plates, where all objects are assumed to be perfectly
reflecting. Using the general expression for the Casimir energy of
multiple objects, Eq.~\eqref{Elogdet}, the energy can be
straightforwardly computed by truncating the matrix $\mathbb{M}$ at a
finite partial wave order $n$. Including up to $n=35$ partial waves,
we obtain for the Casimir force between two cylinders of equal radii
in the presence of one or two sidewalls the results shown in
Fig.~\ref{fig:2cyl+plates}. In this figure the force at a fixed
surface-to-surface distance $d-2R=2R$ between the cylinders is plotted
as a function of the relative separation $(H-R)/R$ between the plate
and cylinder surfaces. Two interesting features can be
observed. First, the attractive total force varies non-monotonically
with $H$: Decreasing for small $H$ and then increasing towards the
asymptotic limit between two isolated cylinders for large $H$,
cf. Eq.~(\ref{eq:E-two-cylinders}). The extremum for the one-sidewall
case occurs at $H-R \approx 0.27 R$, and for the two-sidewall case is
at $H-R \approx 0.46 R$. Second, the total force for the two-sidewall
case in the proximity limit $H=R$ is larger than for $H/R
\rightarrow\infty$. As might be expected, the $H$-dependence for one
sidewall is weaker than for two sidewalls, and the effects of the two
sidewalls are not additive: not only is the difference from the
$H\rightarrow\infty$ force not doubled for two sidewalls compared to
one, but the two curves actually intersect at a separation of
$H/R=1.13$.  The non-monotonic sidewall effect arises from a
competition between the force from TE and TM modes as demonstrated by
the results in Fig.~\ref{fig:2cyl+plates}.  The qualitatively 
different behavior of TE and TM modes can be understood intuitively on
the basis of the method of images\cite{Rahi08-2}.
The non-monotonicity in $H$ also implies that
the force between the cylinders and the sidewalls is not monotonic in
$d$ \cite{Rahi08-2}.

Second, we replace the two cylinders by two identical, general
polarizable compact objects that we specialize later on to spheres
\cite{Rodriguez09}. The meaning of the lengths $d$ and $H$ remains
unchanged.  In dipole approximation, the retarded limit of the
interaction is described by the static electric ($\alpha_z$,
$\alpha_\|$) and magnetic ($\beta_z$, $\beta_\|$) dipole
polarizabilities of the objects which can be different in the
directions perpendicular ($z$) and parallel ($\|$) to the wall.  The
well-known Casimir-Polder (CP) potential between two compact objects
at large distance is
\begin{equation}
  \label{eq:E_CP}
{ \mathcal E}_{2,|}(d) = -\frac{\hbar c}{8\pi d^7} \!\!
\left[ 33 \alpha_\|^2 +\! 13 \alpha_z^2
- \! 14 \alpha_\|\beta_z + (\alpha \!\leftrightarrow\! \beta) \!\right] \, .
\end{equation}
When a sidewall is adde  , the energy changes. Its 
$d$-dependent part is then
\begin{equation}
  \label{eq:E_CP_plane}
  {\mathcal E}_{\underline{\circ\circ}}(d,H) = {\mathcal E}_{2,|}(d) +
  {\mathcal E}_{2,\backslash}(D,d) + {\mathcal E}_3(D,d) 
\end{equation}
with $D=\sqrt{d^2+4H^2}$. The change in the relative orientation of the
objects with $\ell=d/D$ leads to a modification of the 2-body CP potential
\begin{equation}
\label{eq:E_CP_diag}
\begin{split}
\raisetag{35pt}
  {\mathcal E}_{2,\backslash}(D,d) &= -\frac{\hbar c}{8\pi D^7} \!\!\left[ 26\alpha_\|^2
+\! 20 \alpha_z^2 -\! 14 \ell^2 (4\alpha_\|^2-9\alpha_\|\alpha_z
+5\alpha_z^2)\right.\\
& + \left. 63\ell^4 (\alpha_\| - \alpha_z)^2  
- 14\!\left(\alpha_\| \beta_\|(1\!-\!\ell^2) +\!\ell^2 \alpha_\| \beta_z \!\right) + (\alpha\!\leftrightarrow \!\beta) \right] \, .
\end{split}
\end{equation}
The three-body energy ${\mathcal E}_3(D,d)$ describes the collective interaction
between the two objects and one image object.  It is given by
\begin{equation}
  \label{eq:E_3}
\begin{split}
\raisetag{15pt}
  {\mathcal E}_3(D,d) &=  \frac{4\hbar c}{\pi} \frac{1}{d^3D^4(\ell+1)^5}\left[ \Big(
3\ell^6 +15\ell^5+28\ell^4+20\ell^3+6\ell^2-5\ell-1\right)\\
&\times \left(\alpha_\|^2-\beta_\|^2\right)
- \left(3\ell^6+15\ell^5+24\ell^4-10\ell^2-5\ell-1\right) 
\left(\alpha_z^2-\beta_z^2\right)\\
& +4\left(\ell^4+5\ell^3+\ell^2\right)\left(\alpha_z\beta_\|-\alpha_\|\beta_z
\right)\Big] \, .
\end{split}
\end{equation}
It is instructive to consider the two limits $H\ll d$ and $H\gg d$.
For $H\ll d$,
${\mathcal E}_{\underline{\circ\circ}}$ turns out to be the CP
potential of Eq.~\eqref{eq:E_CP} with the replacements $\alpha_z\to
2\alpha_z$, $\alpha_\|\to 0$, $\beta_z\to 0$, $\beta_\|\to
2\beta_\|$. The two-body and three-body contributions add constructively or
destructively, depending on the relative orientation of a dipole and
its image which together form a dipole of zero or twice the original
strength \cite{Rodriguez09}.

For $H \gg d$ the leading correction to the CP potential of
Eq.~\eqref{eq:E_CP} comes from the three-body energy. The energy then
becomes (up to order $H^{-6}$)
\begin{equation}
  \label{eq:E_3_large_H}
  {\mathcal E}_{\underline{\circ\circ}}(d,H) = { \mathcal E}_{2,|}(d)+\frac{\hbar c}{\pi} \!\!\left[ \!\frac{\alpha_z^2-\alpha_\|^2}{4 d^3H^4}  +
\frac{9\alpha_\|^2-\alpha_z^2 -2\alpha_\| \beta_z}{8dH^6} - (\alpha\leftrightarrow \beta)\!\right] .
\end{equation} 
The signs of the polarizabilities in the leading term $\sim H^{-4}$
can be understood from the relative orientation of the dipole of one
object and the image dipole of the other object \cite{Rodriguez09}.

Next, we study the case where the two objects are perfectly reflecting
spheres of radius $R$.  Now we consider arbitrary distances and
include higher order multipole contributions.  For $R \ll d,\, H$ and
arbitrary $H/d$ the result for the force can be written as
\begin{equation}
  \label{eq:force-of-L}
  F  = \frac{\hbar c}{\pi R^2} \sum_{j=6}^\infty  f_j(H/d) \left(\frac{R}{d}\right)^{j+2} \, .
\end{equation}
The functions $f_j$ can be computed exactly and their full form is
given for $j=6,\, 7, \, 8$ in Ref.~\cite{Rodriguez09}.
For $H \gg d$ one has $f_6(h) = -1001/16 +3/(4h^6)+ {\cal O}(h^{-8})$,
$f_8(h)=-71523/160+39/(80h^6)+ {\cal O}(h^{-8})$ so that the wall
induces weak repulsive corrections. For $H \ll d$, $f_6(h)=-791/8+6741
h^2/8 +{\cal O}(h^4)$, $f_8(h)=-60939/80 + 582879 h^2/80 +{\cal
  O}(h^4)$ so that the force amplitude decreases when the spheres are
moved a small distance away from the wall. This proves the existence
of a minimum in the force amplitude as a function of $H$ for fixed,
sufficiently small $R/d$. 

To obtain the interaction at smaller separations or larger radius, the
energy ${\mathcal E}_{\underline{\circ\circ}}$ and force $F=-\partial
{\mathcal E}_{\underline{\circ\circ}} /\partial d$ between the spheres
has been computed numerically \cite{Rodriguez09}.  In order to show
the effect of the sidewall, the energy and force between the spheres,
normalized to the results for two spheres without a wall, is shown in
Fig.~\ref{fig:2spheres+plate} for fixed $d$. When the spheres
approach the wall, the force first decreases slightly if $R/d \lesssim
0.3$ and then increases strongly under a further reduction of $H$. For
$R/d \gtrsim 0.3$ the force increases monotonically as the spheres
approach the wall.  This agrees with the prediction of the large
distance expansion. The expansion of Eq.~\eqref{eq:force-of-L} with
$j=10$ terms is also shown in Fig.~\ref{fig:2spheres+plate} for
$R/d\le 0.2$. Its validity is limited to large $d/R$ and not too small
$H/R$; it fails completely for $R/d>0.2$ and hence is not shown in
this range.

\subsection{Orientation dependence}
\label{subsec:Orientation}

\begin{figure}[ht]
\sidecaption[t]
\includegraphics[width=0.64\linewidth]{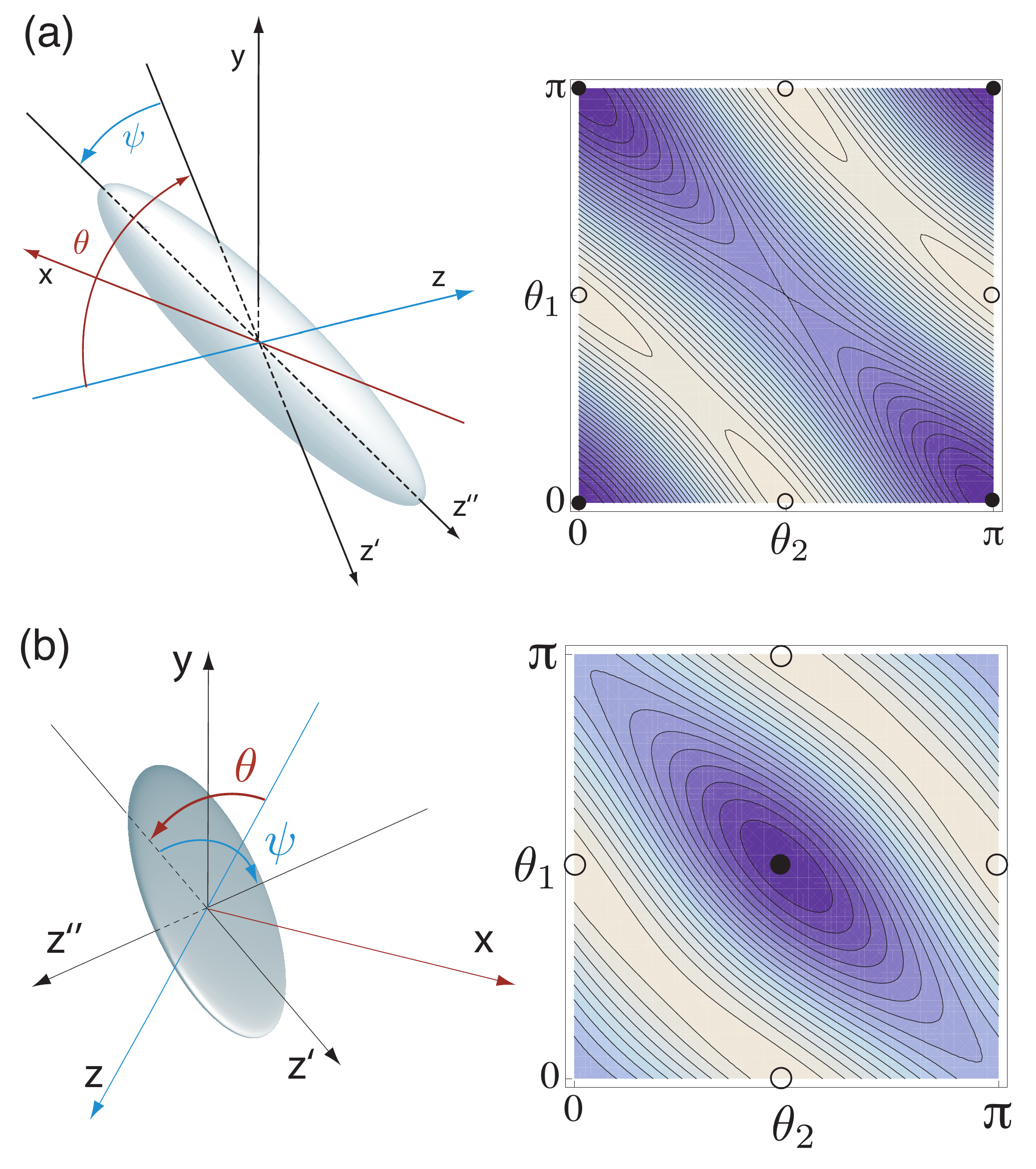}
\caption{
(a) Orientation of a prolate (cigar-shaped)
  spheroid: The symmetry axis (initially the $z$-axis) is rotated by
  $\theta$ about the $x$-axis and then by $\psi$ about the $z$-axis.
  For two such spheroids, the energy at
  large distances is give by Eq.~\eqref{eq:energy-cylidenr-general}.
 The latter is depicted at fixed distance $d$, and for
  $\psi_1=\psi_2$, by a contour plot as function
  of the angles $\theta_1$, $\theta_2$ for the $x$-axis rotations . 
  Minima (maxima) are marked by filled (open) dots.
(b) As in (a) for oblate (pancake-shaped) spheroids, with a
  contour plot of energy at large separations.
}
\label{fig:spheroids}
\end{figure}

In this {subsection} we describe the shape and orientation dependence of
the Casimir force using Eq.~\eqref{Elogdet}, first reported in Ref.~\cite{Emig09}. We consider the
orientation dependent force between two spheroids, and between a
spheroid and a plane.  For two anisotropic objects,
the CP potential of Eq.~\eqref{eq:E_CP} must be generalized.  In terms
of the Cartesian components of the standard electric (magnetic)
polarizability matrix $\mathbb{\alpha}$ ($\mathbb{\beta}$), the
asymptotic large distance potential of two objects (with the $\hat{z}$
axis pointing from one object to the other), can be written as
\begin{equation}
 \label{eq:energy_aniso}
\begin{split}
 {\mathcal E} &=  -\frac{\hbar c}{d^7} \frac{1}{8\pi} \bigg\{
13\left( \alpha^1_{xx}\alpha^2_{xx} + \alpha^1_{yy}\alpha^2_{yy}+2 \alpha^1_{xy}\alpha^2_{xy}\right) \\
&+ 20 \, \alpha^1_{zz}\alpha^2_{zz} -30 \left( \alpha^1_{xz}\alpha^2_{xz} 
+ \alpha^1_{yz}\alpha^2_{yz}\right) +
\left(\mathbb{\alpha}\to\mathbb{\beta}\right) \\
&- 7 \left( \alpha^1_{xx}\beta^2_{yy} +  \alpha^1_{yy}\beta^2_{xx} 
-2 \alpha^1_{xy}\beta^2_{xy} \right) +\left( 1\leftrightarrow 2\right)
\bigg\} \, .
\end{split}
\end{equation} 
 For the case of an
ellipsoidal object with static electric permittivity $\epsilon$ and
magnetic permeability $\mu$, the polarizability tensors are diagonal
in a basis oriented to its principal axes, with elements (for
$i\in\{1,2,3\}$)
\begin{equation}
\label{eq:pol-tensor-diag}
\alpha_{ii}^0 = \frac{V}{4\pi} \frac{\epsilon-1}{1+(\epsilon-1)n_i}\, ,\,
\beta_{ii}^0 = \frac{V}{4\pi} \frac{\mu-1}{1+(\mu-1)n_i}\,,
\end{equation}
where $V=4\pi r_1 r_2 r_3/3$ is the ellipsoid's volume. In the case of
spheroids, for which $r_1=r_2=R$ and $r_3 = L/2$, the so-called
depolarizing factors, $n_{j}$, can be expressed in terms of elementary
functions, $n_1=n_2=\frac{1-n_3}{2}, \, n_3 = \frac{1-e^2}{2e^3} (\log
\frac{1+e}{1-e} - 2 e )$, where the eccentricity $e = (1 -
\frac{4R^2}{L^2})^{1/2}$ is real for a prolate spheroid ($L > 2R$) and
imaginary for an oblate spheroid ($L < 2R$). The polarizability
tensors for an arbitrary orientation are then obtained as
$\mathbb{\alpha}={\cal R}^{-1}\mathbb{\alpha}^0{\cal R}$, where ${\cal
  R}$ is the matrix that orients the principal axis of the spheroid relative
to a fixed Cartesian basis.
Note that for
rarefied media with $\epsilon\simeq 1$, $\mu\simeq 1$ the
polarizabilities are isotropic and proportional to the volume.  Hence,
to leading order in $\epsilon-1$ the interaction is orientation
independent at asymptotically large separations, as we would expect,
since pairwise summation is valid for $\epsilon-1\ll 1$. In the
following we focus on the interesting opposite limit of two identical
perfectly reflecting spheroids. We first consider prolate spheroids
with $L \gg R$.  The orientation of each ``needle'' relative to the
line joining them (the initial $z$-axis) is parameterized by the two
angles $(\theta,\psi)$, as depicted in Fig.~\ref{fig:spheroids}(a).
Then the energy is
\begin{equation}
\label{eq:energy-cylidenr-general}
\begin{split}
{\cal E}(\theta_1,\theta_2,\psi) &= -\frac{\hbar c}{d^7} \bigg\{
\frac{5L^6}{1152 \pi \left( \ln \frac{L}{R} - 1\right)^2}
\bigg[\cos^2\theta_1 \cos^2\theta_2\\
+ &\frac{13}{20}\cos^2\psi \sin^2 \theta_1\sin^2\theta_2
- \frac{3}{8} \cos\psi \sin 2\theta_1 \sin 2\theta_2\bigg]
+{\cal O}\bigg(\frac{L^4R^2}{\ln\frac{L}{R}}\bigg)\bigg\}\, ,
\end{split}
\end{equation}
where $\psi\equiv\psi_1-\psi_2$.  It is minimized for two
needles aligned parallel to their separation vector.  At almost all
orientations the energy scales as $L^6$, and vanishes logarithmically
slowly as $R\to 0$.  The latter scaling changes when one needle is
orthogonal to $\hat{z}$ (i.e. $\theta_1=\pi/2$), while the other is
either parallel to $\hat{z}$ ($\theta_2=0$) or has an arbitrary
$\theta_2$ but differs by an angle $\pi/2$ in its rotation about the
$z$-axis (i.e. $\psi_1-\psi_2=\pi/2$).  In these cases the energy
comes from the next order term in
Eq.~(\ref{eq:energy-cylidenr-general}), and takes the form
\begin{equation}
  \label{eq:crossed-cigars-finite-theta}
  {\cal E}\left(\frac{\pi}{2},\theta_2,\frac{\pi}{2}\right) = 
-\frac{\hbar c}{1152 \pi \, d^7} \frac{L^4R^2}{\ln\frac{L}{R} - 1} 
\left( 73+7\cos 2\theta_2
  \right) \, ,
\end{equation}
which shows that the least favorable configuration corresponds to two
needles orthogonal to each other and to the line joining them.

For perfectly reflecting oblate spheroids with $R\gg L/2$, the
orientation of each ``pancake'' is again described by a pair of angles
$(\theta,\psi)$, as depicted in Fig.~\ref{fig:spheroids}(b). To leading
order at large separations, the energy is given by
\begin{equation}
  \label{eq:energy_oblate}
\begin{split}
  {\mathcal E} &= -\frac{\hbar c}{d^7} \bigg\{
\frac{R^6}{144\pi^3} \bigg[
765 - 5(\cos 2\theta_1+\cos 2\theta_2) 
+237 \cos 2\theta_1 \cos 2\theta_2 \\
&+372 \cos 2\psi \sin^2\theta_1\sin^2\theta_2 
- 300 \cos \psi\sin 2\theta_1 \sin 2\theta_2 \bigg] 
+{\cal O}\big( {R^5L}\big)\bigg\} \, .
\end{split}
\end{equation}
The leading dependence is proportional to $R^6$, and does not
disappear for any choice of orientations.  Furthermore, this
dependence remains even as the thickness of the pancake is taken to
zero ($L\to 0$). This is very different from the case of the needles,
where the interaction energy vanishes with thickness as
$\ln^{-1}(L/R)$.  The lack of $L$ dependence is due to the assumed
perfectly reflectivity.  The energy is minimal for two pancakes
lying on the same plane ($\theta_1=\theta_2=\pi/2$, $\psi=0$) and has
energy $-\hbar c \, (173/18\pi^3) R^6/d^7$.  When the two pancakes are
stacked on top of each other, the energy is increased to $-\hbar c
\,(62/9\pi^3) R^6/d^7$.  The least favorable configuration is when the
pancakes lie in perpendicular planes, i.e., $\theta_1=\pi/2$,
$\theta_2=0$, with an energy $-\hbar c\, (11/3\pi^3) R^6/d^7$.

For an anisotropic object interacting with a perfectly reflecting
mirror, at leading order the CP potential is given by
\begin{equation}
\label{eq:energy_aniso_wall}
{\mathcal E} = -\frac{\hbar c}{d^4} \frac{1}{8\pi} \tr (\alpha-\beta ) 
+{\cal O}(d^{-5})\, ,
\end{equation}
which is clearly independent of orientation.  Orientation dependence
in this system thus comes from higher multipoles.  The next order also
vanishes, so the leading term is the contribution from the partial
waves with $l=3$ for which the scattering matrix is not known
analytically.  However, we can obtain the preferred orientation by
considering a distorted sphere in which the radius $R$ is deformed to
$R+\delta f(\vartheta,\varphi)$.  The function $f$ can be expanded
into spherical harmonics $Y_{lm}(\vartheta,\varphi)$, and spheroidal
symmetry can be mimicked by choosing $f=Y_{20}(\vartheta,\varphi)$.
The leading orientation dependent part of the energy is then obtained
as
\begin{equation}
{\mathcal E}_f = - \hbar c \frac{1607}{640 \sqrt{5} \pi^{3/2}} \frac{\delta R^4}{d^6} \cos(2\theta)   \,. 
\end{equation}
A prolate spheroid ($\delta>0$) thus minimizes its energy by pointing
towards the mirror, while an oblate spheroid ($\delta<0$) prefers to
lie in a plane perpendicular to the mirror.  (It is assumed that the
perturbative results are not changed for large distortions.)  These
configurations are also preferred at small distances $d$, since (at
fixed distance to the center) the object reorients to minimize the
closest separation.  Interestingly, the latter conclusion is not
generally true. In Ref.~\cite{Emig09} it has been shown that there
can be a transition in preferred orientation as a function of $d$ in
the simpler case of a scalar field with Neumann boundary conditions.
The separation at which this transition occurs varies with the
spheroid's eccentricity.

\subsection{Edge and finite size effects}
\label{subsec:Edge}

In this subsection, based on work reported 
in Ref.~\cite{Graham:2010kx}, it is demonstrated that \emph{parabolic} cylinders
provide another example were the scattering amplitudes can be computed
exactly.  We use the exact results for scattering from perfect mirrors
to compute the Casimir force between a parabolic cylinder and a plate.
In the limiting case when the curvature at its tip vanishes, the
parabolic cylinder becomes a semi-infinite plate (a knife's edge), and
we can consider how edges and finite size effects influence the
Casimir energy.

\begin{figure}[ht]
\sidecaption[t]
\includegraphics[width=0.64\linewidth]{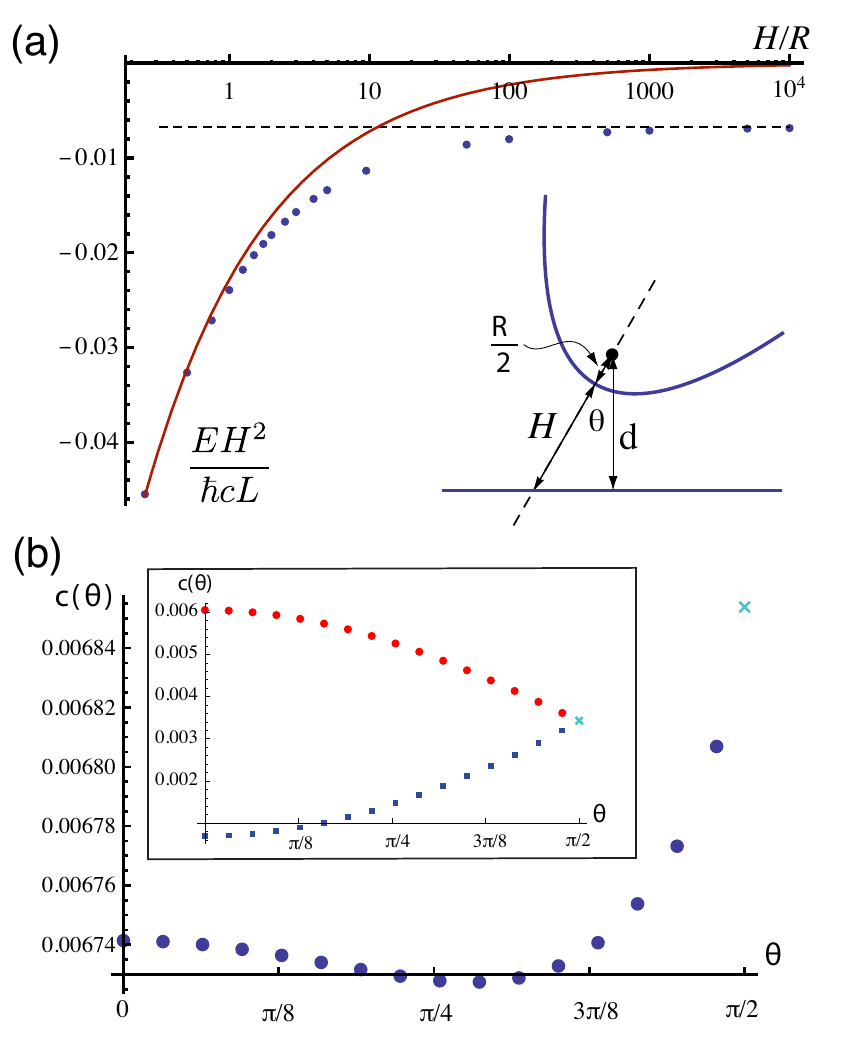}
\caption{ (a) Energy $EH^2/(\hbar c L)$ versus $H/R$ for $\theta=0$
  and $R=1$ on a log-linear scale for the parabolic cylinder-plane
  geometry.  The dashed line gives the $R=0$ limit
  and the solid curve gives the PFA result.
  (b) The coefficient $c(\theta)$ as a function of angle for $R=0$.
  The exact result at $\theta=\pi/2$ is marked with a cross.  Inset:
  Dirichlet (circles) and Neumann (squares) contributions to the full
  electromagnetic result.  }
\label{fig:parabolic_cyl}
\end{figure}

The surface of a parabolic cylinder in Cartesian coordinates is
described by $y=(x^2-R^2)/2R$ for all $z$, as shown in
Fig.~\ref{fig:parabolic_cyl}(a), where $R$ is the curvature at the
tip.  In parabolic cylinder coordinates, defined through
$x=\mu\lambda$, $y=(\lambda^2-\mu^2)/2$, $z=z$, the surface is simply
$\mu=\mu_0=\sqrt{R}$ for $-\infty<\lambda,z<\infty$.
Since sending $\lambda \to -\lambda$ and
$\mu \to -\mu$ returns us to the same point, we restrict our attention
to $\mu\ge 0$ while considering all values of $\lambda$.  Then $\mu$
plays the role of the ``radial'' coordinate in scattering theory and
one can again define regular and outgoing waves \cite{Graham:2010kx}. 
Since both objects are perfect mirrors, translational symmetry along the
$z$-axis enables us to decompose the electromagnetic field into two
scalar fields, as in the case of circular cylinders in Subsection \ref{subsec:1D}.
Each scalar field, describing $E$ (Dirichlet) or $M$ (Neumann) modes,
can then be treated independently, with the sum of their contributions
giving the full electromagnetic result. 

The scattering amplitude of the plate is expressed in a plane wave
basis and is given by Eq.~\eqref{Tplate} with $r^M=-1$ and $r^E=1$.
The scattering amplitude of the parabolic cylinder for $E$ and $M$
polarization is obtained in a parabolic cylinder wave basis as
\cite{Graham:2010kx}
\begin{equation}
\label{eq:F_parabolic}
\begin{split}
{\cal F}_{\rm{para},k_z \nu E,k_z' \nu' E}^{ee}  &= -
\frac{2\pi}{L}\delta(k_z-k_z') \delta_{\nu,\nu'} f_{k_z \nu E}, \quad
f_{k_z \nu E}=i^{\nu}  \frac{D_{\nu}(i\tilde \mu_0)}{D_{-\nu-1}(\tilde \mu_0)}\\
{\cal F}_{\rm{para},k_z \nu M,k_z' \nu' M}^{ee} &= -
\frac{2\pi}{L}\delta(k_z-k_z') \delta_{\nu,\nu'}  f_{k_z \nu M}, \quad 
f_{k_z \nu M} =i^{\nu+1} 
\frac{D_{\nu}'(i\tilde \mu_0)}{D_{-\nu-1}'(\tilde \mu_0)} \, ,
\end{split}
\end{equation}
with $\tilde \mu_0=\sqrt{2R\sqrt{\kappa^2+k_z^2}}$ and the parabolic
cylinder function $D_\nu(u)$ for integer $\nu$. 

For the present geometry, the general formula for the Casimir energy
per unit length can be expressed explicitly as
\begin{equation}
\frac{\cal E}{\hbar c L} = \int_0^\infty
\frac {d\kappa}{2 \pi}
\int_{-\infty}^\infty \frac {dk_z}{2 \pi} 
\log \det \left(\delta_{\nu, \nu'} - 
f_{k_z \nu P} \int_{-\infty}^\infty \hspace{-8pt} d k_x\,
{\cal U}_{\nu k_x k_z}(d,\theta)
r^P {\cal U}_{\nu' k_x k_z} (d,-\theta)
\right) 
\end{equation}
for polarization $P=E$ or $M$.  Here the matrix ${\cal U}$ with
elements
\begin{equation}
{\cal U}_{\nu k_x k_z}(d,\theta)=
\sqrt{\frac{i}{2 k_y \nu! \sqrt{2\pi}}}
\frac{\left(\tan \frac{\phi + \theta}{2}\right)^{\nu}}
{\cos \frac{\phi + \theta}{2}} e^{i k_y d} 
\end{equation}
with $k_y = i\sqrt{\kappa^2+k_x^2+k_z^2}$ and $\tan\phi=k_x/k_y$
describes the translation from parabolic cylinder to plane waves over
the distance $d$ from the focus of the parabola to the plane where
$\theta$ is the angle of inclination of the parabolic cylinder.

Numerical computations of the energy are performed by truncating the
determinant at index $\nu_{\hbox{\tiny max}}$.  For the numbers quoted
below, we have computed for $\nu_{\hbox{\tiny max}}$ up to $200$ and
then extrapolated the result for $\nu_{\hbox{\tiny max}} \to \infty$,
and in Fig.~\ref{fig:parabolic_cyl} we have generally used
$\nu_{\hbox{\tiny max}} = 100$. The dependence of the energy on the
separation $H = d-R/2$ for $\theta=0$ is shown in
Fig.~\ref{fig:parabolic_cyl}(a). At small separations
($H/R\ll 1$) the proximity force approximation, given by
\begin{equation}
\label{eq:pfa}
\frac{{\cal E}_{\hbox{\tiny pfa}}}{\hbar c L} = 
-\frac{\pi^2}{720} \int_{-\infty}^\infty
\frac{dx}{\left[H + x^2/(2R)\right]^3} = -\frac{\pi^3}{960\sqrt{2}}
\sqrt{\frac{R}{H^5}}\,,
\end{equation}
should be valid.  The numerical results in Fig.~\ref{fig:parabolic_cyl}(a) indeed
confirm this expectation.

A more interesting limit is obtained when $R/H\to 0$, corresponding to
a semi-infinite plate for which the PFA energy vanishes.  The exact
result for the energy for $R=0$ and $\theta=0$ is
\begin{equation}
\label{eq:zeroR}
\frac{{\cal E}}{\hbar c L} = -\frac{C_\perp}{H^2}, 
\end{equation}
where $C_\perp=0.0067415$ is obtained by numerical integration.  When
the semi-infinite plate is tilted by an angle $\theta$, dimensional
analysis suggest for the Casimir energy \cite{Gies:2006fk,Weber:2009uq}
\begin{equation}
\frac{{\cal E}}{\hbar c L} = -\frac{C(\theta)}{H^2} \,.
\label{eq:Etheta}
\end{equation}
The function $c(\theta)=\cos(\theta)C(\theta)$ is shown in
Fig.~\ref{fig:parabolic_cyl}(b).  A particularly interesting limit is
$\theta\to \pi/2$, when the two plates are parallel.  In this case,
the leading contribution to the Casimir energy should be proportional
to the area of the half-plane according to the parallel plate formula,
$E_\parallel /(\hbar cA)= -c_\parallel/H^3$ with
$c_\parallel=\pi^2/720$, plus a subleading correction due to the edge.
Multiplying by $\cos\theta$ removes the divergence in the amplitude
$C(\theta)$ as $\theta\to \pi/2$.  As in \cite{Gies:2006fk}, we assume
$c(\theta\to \pi/2)=c_\parallel/2+ \left(\theta - \pi/2\right)
c_{\hbox{\tiny edge}}$, although we cannot rule out the possibility of
additional non-analytic forms, such as logarithmic or other
singularities.  With this assumption, we can estimate the edge
correction $c_{\hbox{\tiny edge}} = 0.0009$ from the data in
Fig.~\ref{fig:parabolic_cyl}(b).  From the inset in
Fig.~\ref{fig:parabolic_cyl}(b), we estimate the Dirichlet and Neumann
contributions to this result to be $c_{\hbox{\tiny edge}}^D = -0.0025$
and $c_{\hbox{\tiny edge}}^N = 0.0034$, respectively. For extensions
to other geometries with edges, inclusion of thermal fluctuations and
experimental implications, see Ref.~\cite{Graham:2010kx}.

 \subsection{Interior configurations}
\label{subsec:Inside}

In this last {subsection} we consider so-called interior configurations where
one object is contained within another that can be also studied with
the methods introduced in
Sect.~\ref{sec:energy-from-field}. Specifically, we obtain the
electrodynamic Casimir interaction of a conducting or dielectric
object inside a perfectly conducting spherical cavity
\cite{Zaheer:2010yq}. In the case where an object, $i$, lies inside a
perfectly conducting cavity, the outer object~$o$, the Casimir energy
of Eq.~\refeq{Elogdet2} becomes
\begin{align}
\mathcal{E} &= \frac{\hbar c}{2 \pi } \int_0^{\infty} d\kappa \log
\frac{\det (\mathcal{I} -  \mathcal{F}^{ii}_{o}  \mathcal{W}^{io}
  \mathcal{F}^{ee}_{i}  \mathcal{V}^{io})}
{\det (\mathcal{I}- \mathcal{F}^{ii}_o  \mathcal{F}^{ee}_i)} \, ,
\label{eq:master}
\end{align}
where $\mathcal{F}^{ii}_{o}$ is the scattering amplitude for interior
scattering of the conducting cavity, a sphere in our case, and
$\mathcal{F}^{ee}_{i}$ the scattering amplitude of the interior
object. The amplitude matrix for interior scattering is the inverse of
the corresponding exterior matrix.  These scattering amplitudes are
evaluated in a spherical vector wave basis with respect to
appropriately chosen origins within each object.  The translation
matrices, $\mathcal{W}^{io}$ and $\mathcal{V}^{io}$, relate regular
wave functions between the coordinate systems of the interior object
and the spherical cavity, see Ref.~\cite{Rahi09} for details.  The
determinant in the denominator of Eq.~(\ref{eq:master}) subtracts the
Casimir energy when the origins of the two objects coincide. This way
of normalizing the Casimir energy differs from the exterior cases
considered before, where the objects are removed to infinite
separation; a choice that would be unnatural in the interior case.

\begin{figure}[ht]
\sidecaption[t]
\includegraphics[width=0.64\linewidth]{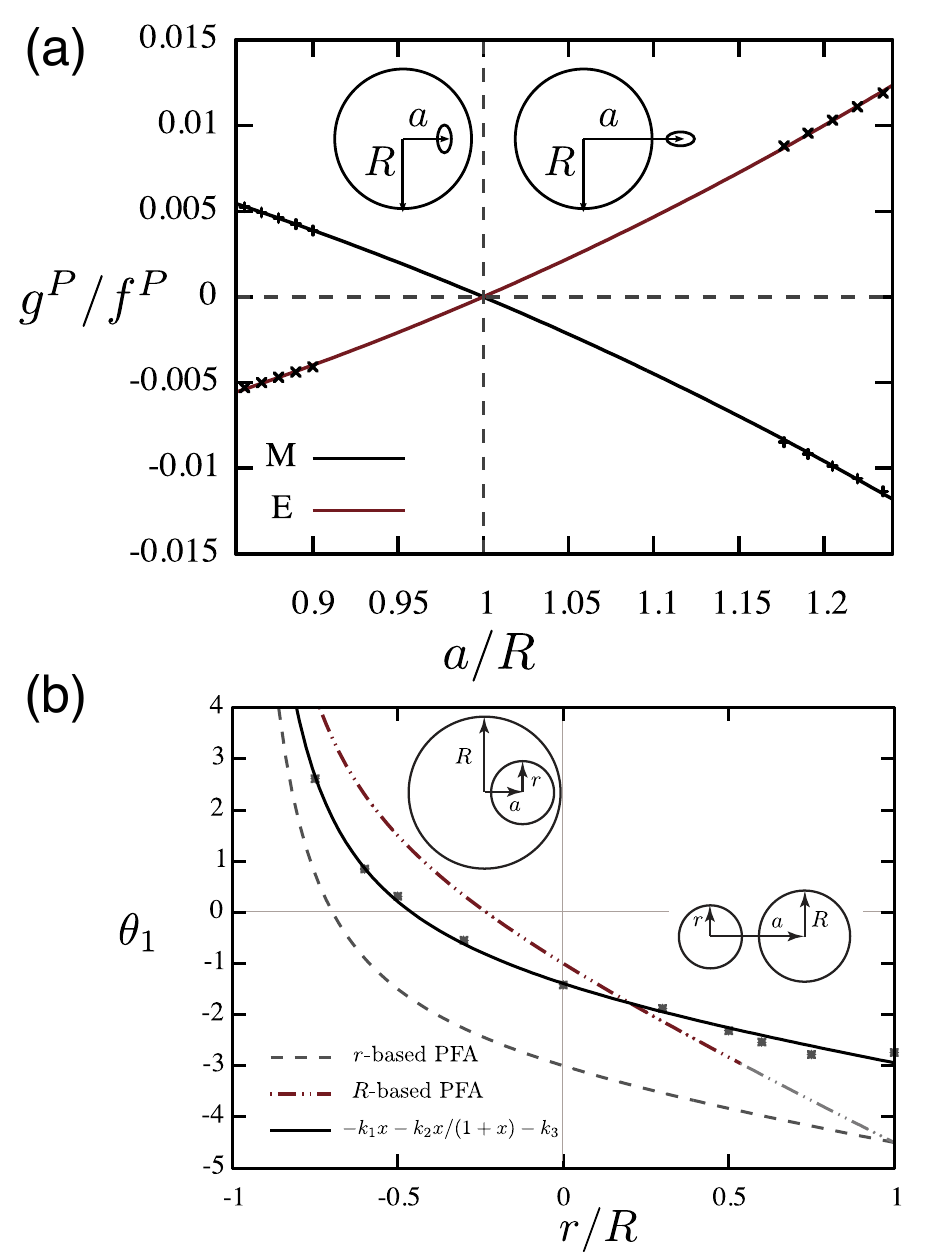}
\caption{(a) The ratio $g^P/f^P$, which determines the preferred
  orientation of the interior object, plotted versus $x=a/R$ showing
  the change in preferred orientation from interior ($a/R<1$) to
  exterior ($a/R>1$) (displayed by two small ellipses as described in
  the text). The solid curves are fits of the form
  $c_1(1-x)+c_2(1-x)^2$ to these data points.  (b) PFA correction
  coefficients for spheres. $r/R$ ranges from $-1$ (interior
  concentric), to zero (sphere-plane), to $+1$ (exterior, equal
  radii).  The data points correspond to the exact values of
  $\theta_1$ calculated numerically, while the solid black curve is a
  fit (see text). Inset: ``interior'' and ``exterior'' geometrical
  configurations.}
\label{fig:int_spheres}
\end{figure}

First, we determine the forces and torques on a small object,
dielectric or conducting, well separated from the cavity walls. This
is the interior analogue of the famous Casimir-Polder force on a
polarizable molecule near a perfectly conducting plate
\cite{Casimir48-1}.  In this case the first term in a multiple
scattering expansion, where the integrand of Eq.~(\ref{eq:master}) is
replaced by $-\text{Tr}(\mathcal{F}^{ii}_{o} \mathcal{W}^{io}
\mathcal{F}^{ee}_{i} \mathcal{V}^{io})$, already gives an excellent
approximation to the energy. Since the object is small, the scattering
amplitude $\mathcal{F}^{ee}_{i,lmP,l'm'P'}$, (where $l$ and $m$ are
angular momentum indices and $P$ labels $M$ or $E$ polarization) can
be expanded in powers of $\kappa$. Only the following terms contribute
to lowest order: $\mathcal{F}^{ee}_{i,1mP,1m'P}(\kappa) =
2\kappa^{3}\alpha^{P}_{mm'}/3 + O(\kappa^4)$, where $\alpha^{P}_{mm'}$
is the static electric ($P=E$) or magnetic ($P=M$) polarizability
tensor of the inner object.  We consider an exterior spherical shell
of radius $R$ and define $a$ to be the displacement of the center of
the interior object from the center of the shell.  Using the dipole
approximation for the inner object but including all multipoles of the
exterior shell, we find for the Casimir energy to leading order in
$r/R$ (where $r$ is the typical length scale of the interior object),
the energy
\begin{equation}
\begin{split}
  \frac{3\pi R^{4}}{\hbar c} \mathcal{E}(a/R)     & = \left[f^{E}(a
    /R)-f^E(0)\right]  {\text{Tr} \, \alpha^E}  \\  
 & + g^{E}(a/R) (2 \alpha^E_{zz}-\alpha^E_{xx}- \alpha^E_{yy})   + (E
 \leftrightarrow M). 
\label{eq:orientation}
\end{split}
\end{equation}
The $z$-axis is oriented from the center of the shell to the innterior
object, and $\alpha^{P}_{ij}$ represent the interior object's static
polarizability tensors in a Cartesian basis. The coefficient functions
$f^P$ and $g^P$ can be obtained in terms of an integral over modified
Bessel functions, see Ref.~\cite{Zaheer:2010yq}.  $f^E$ is negative
and decreasing with $a/R$, while $f^M$ is positive and increasing.
There are important differences between Eq.~(\ref{eq:orientation}) and
the classic Casimir-Polder result: first, the energy depends in a
non-trivial way on $a/R$; second, at any non-zero distance from the
center, the interior object experiences a torque; and third, the force
between the two bodies depends on the interior object's orientation.

To explore the orientation dependence of Eq.~(\ref{eq:orientation})
assume, for simplicity, there is a single frame in which both
$\alpha^{E}$ and $\alpha^{M}$ are diagonal. In this body-fixed frame,
write $\alpha^{0}_{xx}-\alpha^{0}_{yy} = \beta$ and
$\alpha^{0}_{zz}-\frac{1}{2}(\alpha^{0}_{xx}+\alpha^{0}_{yy}) =
\gamma$ (where we have suppressed the $M/E$ label). The polarizability
in the ``lab frame'' is obtained by $\alpha =\mathcal{R} \alpha^{0}
\mathcal{R}^{-1}$, where $\mathcal{R}$ is a rotation matrix that
orients the principal axes of the inner object with respect to the lab
frame. This procedure leaves $\text{Tr}\alpha^0$ invariant, and gives
for the second line in Eq.~(\ref{eq:orientation}),
\begin{align}
 \sum_{P = M,E}g^{P}(a/R)& \left(\frac{3\beta^{P}}{2} \sin^{2}\theta\,\cos 2\phi +\gamma^P(3\cos^{2}\theta-1)\right) \, ,
\nonumber
\end{align}
where $\phi$ corresponds to the azimuthal rotation of the object about
its principal $z$-axis, and $\theta$ is the angle between the object's
principal $z$-axis and the ``laboratory'' $z$-axis connecting the
center of the sphere to the origin of the interior object.
 
If $\beta \ne 0$ then the object held at fixed inclination, $\theta$,
experiences a torque that causes it to rotate about the body-fixed
$z$-axis.  If, however, the object has axial symmetry $(\beta=0$),
then the only torque on the object tries to align it either parallel
or perpendicular to the displacement axis.

A ``cigar shaped'' object ($\gamma>0$) prefers to orient so as to
point perpendicular to the $z$ axis, and a ``pancake'' ($\gamma <0$)
tries to align its two large axes perpendicular to the $z$ axis. The
small ellipse inside the sphere in Fig.~\ref{fig:int_spheres}(a)
illustrates a side view of both the cigar and the pancake in their
preferred orientation. It is interesting to note that $g^E$ and $g^M$
are both positive. So, in contrast to the force, the contributions to
the torque from magnetic and electric polarizabilities are in the same
direction, if they have the same sign. More complicated behavior is
possible if, for example, the electric and magnetic polarizabilities
are not diagonal in the same body-fixed coordinate system. Note that
our results cannot be compared to the PFA approximation since the the
size of the inner object, not the separation of surfaces, $d$, has
been assumed to be the smallest scale in the analysis.

An identical analysis can be performed for a polarizable object
outside a metallic sphere where $a/R>1$. It turns out that the
analogous exterior function $g(a/R) <0$ for both
polarizations. Therefore, the preferred orientation of a polarizable
object outside a metallic sphere is opposite of that in the interior
case (see the small ellipse outside the large sphere in
Fig.~\ref{fig:int_spheres}(a)). The continuation of the functions $f$
and $g$ from ``interior'' to ``exterior'' is displayed in
Fig.~\ref{fig:int_spheres}(a), where the transition from one
orientation to the other is clear.

Second, we compute numerically from Eq.~(\ref{eq:master}) the
interaction energy of a finite-size metal sphere with the cavity walls
when the separation, $d$, between their surfaces tends to zero.  In
this limit the Casimir force $F$ between two conducting spheres, which
is attractive, is proportional in magnitude to $d^{-3}$, where $d=
R-r-a$ is the separation of surfaces. The coefficient of $d^{-3}$ is
given by the PFA,
\begin{align}
\lim_{d\to 0}d^{3}\, F= - \frac{\pi^3 \hbar c}{360} \frac{rR}{r+R} \, . 
\label{pfalimit}
\end{align}
This result holds for both the interior and the exterior configuration
of two spheres. For fixed $r$ we formally distinguish the cases: $R>0$
for the exterior, $R\to\infty$ for the plate-sphere, and $R<0$ for the
interior configuration, see Fig.~\ref{fig:int_spheres}(b). All
possible configurations are taken into account by considering $-1\le
r/R\le 1$. Although we know of no derivation of the functional form of
the Casimir force beyond the leading term in the PFA, our numerical
results are well fit by a power series in $d/r$,
\begin{align}
F= - \frac{\pi^3 \hbar c}{360 d^3 } \frac{rR}{r+R}\left( 1 + \theta_{1}(r/R)\frac{d}{2r} - \theta_{2}(r/R)\frac{d^{2}}{2r^{2}}+... \right)
\end{align}
We have used this functional form to extract the coefficient $\theta_{1}(r/R)$.

Although the PFA is accurate only in the limit $d/r\to 0$, it can be
extended in various ways to the whole range of $d$, $r$, and $R$.
Depending on the surface $O$ from which the normal distance to the
other surface is measured, one obtains the ``$O$-based'' PFA
energy. Clearly, the result depends on which object one chooses as
$O$, but the various results do agree to leading order in $d/r$. We
can choose either of the two spheres to arrive at the ``$r$-based
PFA'' or the ``$R$-based PFA'', see
Fig.~\ref{fig:int_spheres}(b). Either one yields a `correction' to the
leading order PFA,
\begin{align*} 
\theta^{\text{PFA}}_{1,r}(x) = -\left(\!\!x+\frac{x}{1+x}+3\!\right),
\quad
\theta^{\text{PFA}}_{1,R} = -\left(\!\!3x +\frac{x}{1+x}+1\!\right),
\end{align*} 
where $x=r/R$.  In Fig.~\ref{fig:int_spheres}(b) we plot the values of
$\theta_{1}$ extracted from a numerical evaluation of the force from
Eq.~\eqref{eq:master} for various values of $r/R<0$. For reference, the
two PFA estimates are also shown. 

The numerical data in Fig.~\ref{fig:int_spheres}(b) show a smooth
transition from the interior to the exterior configuration. Although
the PFA estimates do not describe the data, the $r$-based PFA has a
similar functional form and divergence as $x\to -1$. Therefore, we fit
the data in Fig.~\ref{fig:int_spheres}(b) to a function, $\theta_1(x)
= -(k_1 x +k_2x/(1+x) +k_3)$ and find, $k_1=1.05 \pm 0.14, k_2 = 1.08
\pm 0.08, k_3 = 1.38 \pm 0.06$. Notice, however, that the actual
function $\theta_1(x)$ is not known analytically and that the fit
represents a reasonable choice which may not be unique. Our results
show that the correction to the PFA has a significant dependence on
ratio of curvatures of the two surfaces.

\begin{acknowledgement}
The research presented here was conducted together with Noah Graham, Steven G. Johnson, Mehran Kardar, Alejandro W. Rodriguez, Pablo Rodriguez-Lopez, Alexander Shpunt, and Saad Zaheer, whom we thank for their collaboration.
  This work was supported by the National Science Foundation (NSF)
  through grant DMR-08-03315 (SJR), by the DFG through grant EM70/3 (TE) and
  by the U. S. Department of Energy (DOE) under cooperative research
  agreement \#DF-FC02-94ER40818 (RLJ).  
\end{acknowledgement}

\input{Springer_20100722.bbl}
%
%
% When placed at the end of a chapter or contribution (as opposed to at the end of the book), the numbering of tables, figures, and equations in the appendix section continues on from that in the main text. Hence please \textit{do not} use the \verb|appendix| command when writing an appendix at the end of your chapter or contribution. If there is only one the appendix is designated ``Appendix'', or ``Appendix 1'', or ``Appendix 2'', etc. if there is more than one.

%\bibliographystyle{spmpsci}
%\bibliography{main_clean}
%%\bibliographystyle{plain}
%\input{biblio}
% \bibliographystyle{}
% \bibliography{}

\end{document}

%% file: chap1.tex
%% This is an example first chapter.  You should put chapter/appendix that you
%% write into a separate file, and add a line \include{yourfilename} to
%% main.tex, where `yourfilename.tex' is the name of the chapter/appendix file.
%% You can process specific files by typing their names in at the 
%% \files=
%% prompt when you run the file main.tex through LaTeX.
\section{Introduction}
\label{sec:Intro}

Neutral objects exert a force on one another through electromagnetic fields even if they do not possess permanent multipole moments.
Materials that couple to the electromagnetic field alter the spectrum
of the field's quantum and thermal fluctuations.  The
resulting change in energy depends on the relative positions of the
objects, leading to a fluctuation-induced force, usually called the
Casimir force. Alternatively, one can regard the cause of these forces to be spontaneous charges and currents, which fluctuate in and out of existence in the objects due to quantum mechanics. The name `Van der Waals force' is sometimes used interchangeably but it usually refers to the Casimir force in the regime where objects are close enough to one another that the speed of light is effectively infinite. The Casimir force has been the subject of precision experimental measurements
\cite{Lamoreaux97,Mohideen98,Roy99,Ederth00,Chan01,Chen02,DeKiewiet03,Harber05,Chen06,Krause07,Decca07,Chen07,Munday07,Chan08,Kim08,Palasantzas08,Munday09}
and can influence the operation of nanoscale devices \cite{Chan01,Capasso07}, see reference \cite{Klimchitskaya09} for a review of the experiments.

Casimir and Polder calculated the fluctuation-induced force on
a polarizable atom in front of a perfectly conducting plate and
between two polarizable atoms, both to leading order at large
separation, and obtained a simple result depending only on the atoms'
static polarizabilities \cite{Casimir48-1}.  Casimir then extended this
result to his famous calculation of the pressure on two perfectly
conducting parallel plates \cite{Casimir48-2}. Feinberg and Sucher
\cite{Feinberg68,Feinberg70}
generalized the result of Casimir and Polder to include both electric
and magnetic polarizabilities.  Lifshitz, Dzyaloshinskii, and
Pitaevskii extended Casimir's result for parallel plates by
incorporating nonzero temperature, permittivity, and permeability into
a general formula for the pressure on two infinite half-spaces
separated by a gap \cite{Lifshitz56,Dzyaloshinskii61,Lifshitz80}. % deleted Lifshitz55,Lifshitz57

\begin{figure}[ht]
\begin{center}
\includegraphics[width=0.5\linewidth]{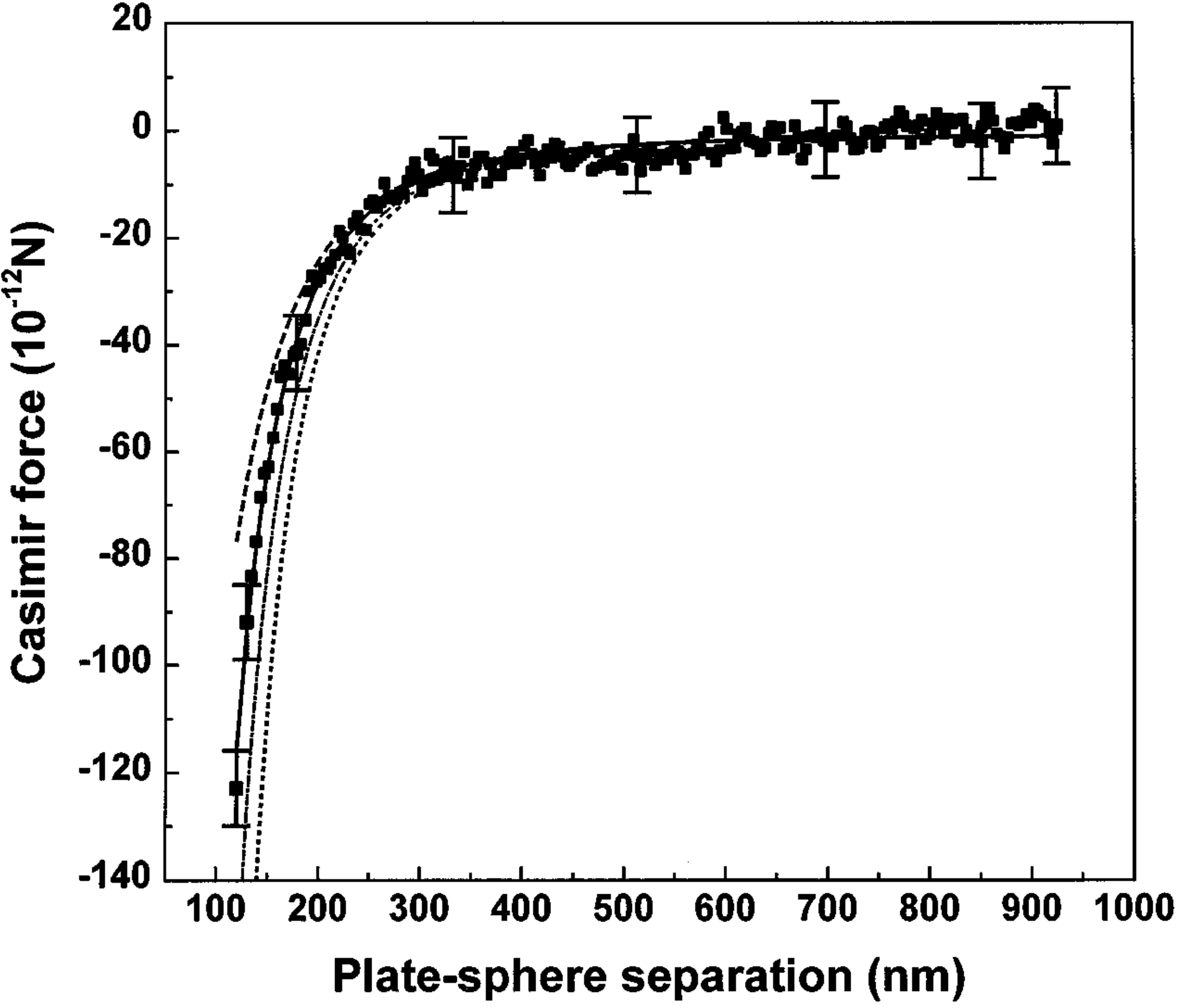}
\caption{
Force between a sphere of radius $\approx 100 \mu\text{m}$ and a plate, both coated with Au-Pd  \cite{Mohideen98}. Square dots represent measurements, the solid line is a theoretical computation using the PFA approximation and taking into account roughness and finite temperature corrections as well as material properties. The other lines represent calculations, where some of these corrections are not taken into account.
}
\end{center}
\label{fig:Mohideen}
\end{figure}

While these early theoretical predictions of the Casimir force applied only to infinite planar geometries (or atoms), the first precision experiments measured the force between a plate and a sphere. This geometry was preferred because keeping two plane surfaces parallel introduces additional challenges for the experimentalist. To compare the measurements with theory, however, a makeshift solution had to be used:  known as the Proximity Force Approximation (PFA), it estimates the Casimir force by integrating the Casimir pressure between opposing infinitesmal surface area elements, as if they were parallel plates, over the area that the sphere and the plate expose to one another~\cite{Parsegian05}. In general, this simple approximation does not capture curvature corrections but in many experimental situations, it performs surprisingly well, as can be seen in \reffig{Mohideen}, for example; at the small separations at which the force is typically probed in precision measurements the sphere and the plate surfaces are well approximated by a collection of infinitesimal parallel plates.

Clearly, for larger separations and for surfaces that are not smooth, the PFA must fail. For example, in measurements of the Casimir force between a sphere and a trench array significant discrepancies were found \cite{Chan08}. And even for the regimes in which the PFA yields good estimates it would be desirable to know what the corrections are.

In order to study Casimir forces in more general geometries, it turns
out to be advantageous to describe the influence of an arrangement of
objects on the electromagnetic field by the way they scatter
electromagnetic waves. Here, we derive and apply a
representation of the Casimir energy, first developed with various limitations in Refs.~\cite{Emig07,Emig08} and then fully generalized in Ref. \cite{Rahi09}, that characterizes each object by its
on-shell electromagnetic scattering amplitude.  The separations and
orientations of the objects are encoded in universal translation
matrices, which describe how a solution to the source-free Maxwell's
equations in the basis appropriate to one object looks when expanded
in the basis appropriate to another.  The translation matrices depend
on the displacement and orientation of coordinate systems, but not on
the nature of the objects themselves.  The scattering amplitudes and
translation matrices are then combined in a simple algorithm that
allows efficient numerical and, in some cases, analytical
calculations of Casimir forces and torques for a wide variety
of geometries, materials, and external conditions.
The formalism applies to a wide variety of circumstances, including:
\begin{itemize}
\item
$n$ arbitrarily shaped objects, whose
surfaces may be smooth or rough or may include edges and cusps;
\item
objects with arbitrary linear electromagnetic response, including
frequency-dependent, lossy electric permittivity and magnetic
permeability tensors;
\item
objects separated by vacuum or by a medium with
uniform, frequency-dependent isotropic permittivity and permeability;
\item
zero or nonzero temperature;
\item
and objects outside of one another or
enclosed in each other.
\end{itemize}

These ideas build on a range of previous related work, an inevitably incomplete subset of which is briefly reviewed here:
Scattering theory methods were first applied to the parallel
plate geometry, when Kats reformulated Lifshitz theory in terms of
reflection coefficients \cite{Kats77}. Jaekel and Reynaud derived the
Lifshitz formula using reflection coefficients for lossless infinite
plates \cite{Jaekel91} and Genet, Lambrecht, and Reynaud
extended this analysis to the lossy case \cite{Genet03}. Lambrecht, Maia
Neto, and Reynaud generalized these results to include non-specular
reflection  \cite{Lambrecht06}.

Around the same time as Kats's work, Balian and Duplantier developed a
multiple scattering approach to the Casimir energy for perfect metal
objects and used it to compute the Casimir energy at asymptotically
large separations \cite{Balian77,Balian78} at both zero and nonzero
temperature. In their approach, information about the conductors is
encoded in a local surface scattering kernel, whose relation to more
conventional scattering formalisms is not transparent, and their approach was
not pursued further at the time. One can find multiple scattering
formulas in an even earlier article by Renne \cite{Renne71}, but
scattering is not explicitly mentioned, and the technique is only used
to rederive older results.

Another scattering-based approach has been to express the Casimir
energy as an integral over the density of states of the fluctuating
field, using the Krein formula \cite{Krein53,Krein62,Birman62} to
relate the density of states to the $S$-matrix for scattering from
the ensemble of objects.  This $S$-matrix is
difficult to compute in general. In studying many-body
scattering, Henseler and Wirzba connected the $S$-matrix of a
collection of spheres \cite{Henseler97} or disks \cite{Wirzba99} to
the objects' individual $S$-matrices, which are easy to
find. Bulgac, Magierski, and Wirzba combined this result with the
Krein formula to investigate the scalar and fermionic Casimir effect
for disks and spheres \cite{Bulgac01,Bulgac06,Wirzba08}. Casimir
energies of solitons in renormalizable quantum field theories
have been computed using scattering theory techniques that combine
analytic and numerical methods \cite{Graham09}.

Bordag, Robaschik, Scharnhorst, and Wieczorek
\cite{Bordag85,Robaschik87} introduced path integral methods to the
study of Casimir effects and used them to investigate the
electromagnetic Casimir effect for two parallel perfect metal plates.
Li and Kardar used similar methods to study the scalar thermal Casimir
effect for Dirichlet, Neumann, and mixed boundary conditions
\cite{Li91,Li92}. The quantum extension was developed further by
Golestanian and Kardar \cite{Golestanian97,Golestanian98} and was
subsequently applied to the quantum electromagnetic Casimir effect
by Emig, Hanke, Golestanian, and Kardar, who studied the Casimir
interaction between plates with roughness \cite{Emig01} and between
deformed plates \cite{Emig03}.  (Techniques developed to study the
scalar Casimir effect can be applied to the electromagnetic case for
perfect metals with translation symmetry in one spatial direction,
since then the electromagnetic problem decomposes into two scalar
ones.)  Finally, the path integral approach was connected to
scattering theory by Emig and Buescher \cite{Buescher05}.

Closely related to the work we present here is that of Kenneth and
Klich, who expressed the data required to characterize
Casimir fluctuations in terms of the transition $\T$-operator for
scattering of the fluctuating field from the objects \cite{Kenneth06}.
Their abstract representation made it possible to prove
general properties of the sign of the Casimir force.
In Refs. \cite{Emig07,Emig08}, we developed a framework in which this
abstract result can be applied to concrete calculations.  In this
approach, the $\T$-operator is related to the 
scattering amplitude for each object individually, which in turn is
expressed in an appropriate basis of multipoles.  
While the $\T$-operator is in general ``off-shell,'' meaning it has
matrix elements between states with different {spatial }frequencies, the
scattering amplitudes are the ``on-shell'' matrix elements of this
operator between states of equal {spatial }frequency.\footnote{Because of this
relationship, these scattering amplitudes are also referred to as
elements of the $T$-\emph{matrix}.  In standard conventions,
however, the $T$-matrix differs from the matrix elements of the
$\T$-operator by a basis-dependent constant, so we will use the term
``scattering amplitude'' to avoid confusion.} {So, it is not the $\T$-operator itself that connects, say, outgoing and standing waves in the case of outside scattering but its on-shell matrix elements, the scattering amplitudes. }In this approach, the
objects can have any shape or material properties, as long as the
scattering amplitude can be computed in a multipole expansion (or
measured).  The approach can be regarded as a
concrete implementation of the proposal emphasized by Schwinger
\cite{Schwinger75} that the fluctuations of the electromagnetic field
can be traced back to charge and current fluctuations on the objects.
This formalism has been applied and extended in a number of Casimir
calculations \cite{Kenneth:2007jk,Milton08-3,Milton08-4,reid-2009,golestanian-2009,Ttira:2009ku}. 

The basis in which the scattering amplitude for each object is
supplied is typically associated with a coordinate system appropriate
to the object.  Of course a plane, a cylinder, or a sphere would be
described in Cartesian, cylindrical, or spherical coordinates, respectively.
However, any compact object can be described, for example, in
spherical coordinates, provided that the matrix of scattering amplitudes
can be either calculated or measured in that coordinate system.
There are a limited number of coordinate systems in which such a
partial wave expansion is possible, namely those for which  the vector
Helmholtz equation is separable. The translation 
matrices for common separable coordinate systems, obtained
from the free Green's function, are supplied in Appendix{~C of reference \cite{Rahi09}.}  For typical cases, the final computation of the
Casimir energy can be performed on a desktop computer for a wide range
of separations. Asymptotic results at large separation can be obtained
analytically.

The primary limitation of the method is on the distance between
objects, since the basis appropriate to a given object may become
impractical as two objects approach.  For small separations, 
sufficient accuracy can only be obtained if the calculation is taken 
to very high partial wave order. {(Vastly different scales are problematic for numerical evaluations in general.)} In the case of two
spheres, the scattering amplitude is available in a spherical basis,
but as the two spheres approach, the Casimir energy is dominated by
waves near the point of closest approach \cite{Schaden:1998zz}.  As
the spheres come into contact an infinite number of spherical waves
are needed to capture the dominant contribution.  A particular basis may also
be fundamentally inappropriate at small separations.  For instance, if
the interaction of two elliptic cylinders is expressed in
an ordinary cylindrical basis, when the elliptic cylinders are close
enough, the smallest circular cylinder enclosing one may not lie outside the smallest circular cylinder enclosing the other. In that case the 
cylindrical basis would not ``resolve'' the two objects (although an
elliptic cylindrical basis would).
Finally, for a variety of conceptual and computational reasons, we are
limited to linear electromagnetic response.

{In spirit and in mathematical form our final result resembles similar expressions obtained in surface integral equation methods used in computational electrodynamics~\cite{Chew01}. Using such a formulation, in which the unknowns are currents and fields on the objects, one can compute the Casimir energy using more general basis functions, e.g., localized basis functions associated with a grid or mesh, giving rise to finite element and boundary elements methods \cite{reid-2009}.}

{In addition to an efficient computational approach, the scattering formalism has provided the basis for proving general theorems regarding Casimir forces. The seemingly natural question whether the force is attractive or repulsive turns out to be an ill-defined or, at least, a tricky one on closer inspection. When, for example, many bodies are considered, the direction of the force on any one object depends, of course, on which other object's perspective one takes. Even for two objects, ``attractive'' forces can be arranged to appear as a ``repulsive'' force, as in the case of two interlocking combs \cite{rodriguez-2008}. To avoid such ambiguous situations one can restrict oneself to analyzing two objects that are separable by a plane. Even here, it has turned out that a simple criterion for the direction of the force could not be found. Based on various calculations for simple geometries it was thought that the direction of the force can be predicted based on the relative permittivities and permeabilities of the objects and the medium.  Separating materials into two groups, with {\em (i)} permittivity higher than the medium or permeability lower than the medium ($\ep>\ep_M$ and $\mu\leq\mu_M$), or {\em (ii)} the other way around ($\ep<\ep_M$ and $\mu\geq\mu_M$), Casimir forces had been found to be attractive between members of the same group and repulsive for different types in the geometries considered. However, a recent counterexample~\cite{Levin10} shows that this is not always true. A rigorous theorem, which states that Casimir forces are always attractive, exists only for the special case of mirror symmetric arrangements of objects. It was proven first with a $\T$-operator formalism \cite{Kenneth06}, similar to our approach used here, and later using reflection positivity \cite{Bachas07}. We have taken an alternative characterization of the force to be fundamental, namely, whether it can produce a stable equilibrium \cite{Rahi09-3}. Here, the categorization of materials into the two groups is meaningful since objects made of materials of the same type cannot produce stable levitation. One practical consequence of this theorem is that it reveals that many current proposals for producing levitation using metamaterials cannot succeed.}

To illustrate the general formulation, we provide some sample
applications{.}
{We include an analysis of the forces between two cylinders or wires \cite{Rahi08-2} and a cylinder and a plate \cite{Emig06,Rahi08-2,Rahi09}. The Casimir interaction of three bodies is presented subsequently; it reveals interesting multibody effects \cite{Rahi08-1,Rahi08-2,Rodriguez09}. The Casimir torque of two spheroids is discussed as well \cite{Emig09}. }
{Furthermore, we analyze the Casimir effect for a parabolic cylinder opposite a plate when both represent perfect metal material boundary conditions \cite{Graham:2010kx}. We find that the Casimir force does not vanish in the limit of an infinitesimally thin parabola, where a half plate is arranged above an infinite plate, and we compute the edge effect.}
{Another type of geometries that is treated here consists of a finite sphere or a small spheroid inside a spherical metallic cavity~\cite{Zaheer:2010yq}.}

{This chapter} is organized as follows: First, we sketch the derivation of the Casimir interaction energy formula Eq. (\ref{Elogdet}) in Section \ref{sec:energy-from-field}.
{Next, the theorem regarding stability is derived in Section \ref{sec:Constraints}.} {Finally, in} Section {\ref{sec:Applications}} sample applications are presented.

%% file: chap2.tex
%%%%%%%%%%%%%%%%%%%%%h%%%%%%%%%%%%%%%%%%%%%%%%%%%%%%%%%%%%%%%%%%%%%%
%%%%%%%%%%%%%%%%%%%%%%%%%%%%%%%%%%%%%%%%%%%%%%%%%%%%%%%%%%%%%%%%%%%
%%%%%%%%%%%%%%%%%%%%%%%%%%%%%%%%%%%%%%%%%%%%%%%%%%%%%%%%%%%%%%%%%%%
\section{General theory for Casimir interactions}
\label{sec:energy-from-field}
\noindent

This section has been adapted from a longer article, reference \cite{Rahi09}. Many technical details and extensive appendices have been omitted to fit the format of this book.

\subsection{Path integral quantization}
\label{sec:PathIntegralQuantization}

%%%%%%%%%%%%%%%%%%%%%%%%%%%%%%%%%%%%%%%%%%%%%%%%%%%%%%%%%%%%%%%%%%%
%%%%%%%%%%%%%%%%%%%%%%%%%%%%%%%%%%%%%%%%%%%%%%%%%%%%%%%%%%%%%%%%%%%
\subsubsection{Electromagnetic Lagrangian}
\label{sec:LandSem}
 
We consider the Casimir effect for objects without free charges and
currents but with nonzero electric and magnetic susceptibilities. The
macroscopic electromagnetic Lagrangian density is
\be
\Lag = \half(\bfE\cdot \bfD-\B\cdot \Hf).
\labeleqn{Lem}
\ee
The electric field $\bfE(t,\vecx)$ and the magnetic field $\B(t,\vecx)$
are related to the fundamental four-vector potential
$A^\mu$ by $\bfE= - c^{-1} \partial_t \A - \bnabla A^0$ and
$\B = \curl \A$. We treat stationary objects whose responses to the
electric and magnetic fields are linear. For such materials, the $\bfD$
and $\B$ fields are related to the $\bfE$ and $\Hf$ fields by the
convolutions $\bfD(t,\vecx) = \int_{-\infty}^\infty dt' \,
\epsilon(t',\vecx) \bfE(t-t',\vecx)$ and $\B(t,\vecx) =
\int_{-\infty}^\infty dt' \, \mu(t',\vecx) \Hf(t-t',\vecx)$ in time, where $\epsilon(t',\vecx)$ and $\mu(t',\vecx)$ vanish for $t'<0$. 
We consider local, isotropic permittivity and permeability, although
our derivation can be adapted to  apply to non-local and non-isotropic
media simply by substituting the appropriate non-local and tensor
permittivity and permeability functions. A more formal derivation of
our starting point \refeqn{Lem}, which elucidates the causality
properties of the permeability and permittivity response functions, is
given in Appendix A of reference \cite{Rahi09}.

We define the quantum-mechanical energy through the path integral,
which sums all configurations of the electromagnetic
fields constrained by periodic boundary conditions in time 
between $0$ and $T$.  Outside of this time interval the fields are
periodically continued.  Substituting the Fourier expansions of the form
$\bfE(t,\vecx) = \sum_{n=-\infty}^\infty \bfE(\omega_n,\vecx) 
e^{-i \omega_n t}$ with $\omega_n = 2\pi n/T$, we obtain the action
\be
S(T) =\half \int_0^T dt\int d\vecx \, \left(\bfE\cdot \bfD-\B\cdot \Hf\right)
= \half T \sum_{n=-\infty}^\infty \int d\vecx
\left(
\bfE^* \cdot  \epsilon \bfE - \B^* \cdot \mu^{-1} \B
\right),
\labeleqn{Sem1}
\ee
where $\epsilon$, $\bfE$, $\mu$, and $\B$ on the right-hand side are
functions of position $\vecx$ and frequency $\omega_n$, and we have
used $\bfD(\omega,\vecx)=\epsilon(\omega,\vecx)\bfE(\omega,\vecx)$ and
$\Hf(\omega,\vecx) = \tfrac{1}{\mu(\omega,\vecx)} \B(\omega,\vecx)$.

From the definition of the fields $\bfE$ and $\B$ in terms of the
vector potential $A^\mu$, we have
$\curl \bfE = i\frac{\omega}{c} \B$, which enables us
to eliminate $\B$ in the action,
\be
S(T) = \half T \sum_{n=-\infty}^\infty \int d\vecx\left[ \bfE^{*} \cdot
\left(\tI - \frac{c^2}{\omega_n^2}
\curl \curl \right) \bfE - \frac{c^2}{\omega_n^2} \bfE^{*} \cdot \tV \, \bfE\right],
\labeleqn{Sem2}
\ee
where
\be
\tV = \tI \, \frac{\omega_n^2}{c^2}
\left(1-\epsilon(\omega_n,\vecx)\right) + \curl
\left(\frac{1}{\mu(\omega_n,\vecx)}-1\right) \curl
\labeleqn{Vem1}
\ee
is the potential operator and we have restored the explicit frequency
dependence of $\epsilon$ and $\mu$.  The potential operator is nonzero
only at those points in space where the objects are located ($\epsilon
\neq 1$ or $\mu \neq 1$).

In the functional integral we will sum over configurations of the
field $A^\mu$.  This sum must be restricted by a
choice of gauge, so that it does not include the infinitely
redundant gauge orbits.  We choose to work in the gauge $A^0=0$,
although of course no physical results depend on this choice.

%%%%%%%%%%%%%%%%%%%%%%%%%%%%%%%%%%%%%%%%%%%%%%%%%%%%%%%%%%%%%%%%%%%
%%%%%%%%%%%%%%%%%%%%%%%%%%%%%%%%%%%%%%%%%%%%%%%%%%%%%%%%%%%%%%%%%%%

\subsubsection{Casimir energy from Euclidean action}
We use standard tools to obtain a functional integral expression for
the ground state energy of a quantum field in a fixed
background described by $\V(\omega,\vecx)$. The overlap between the
initial state $\ket{\bfE_a}$ of a system with the state $\ket{\bfE_b}$ after
time $T$ can be expressed as a functional integral with the fields
fixed at the temporal boundaries~\cite{Feynman65},
\be
\bra{\bfE_b} e^{-i \Ham T\hbar} \ket{\bfE_a} =\int \left.\dA\,
\right|_{^{\bfE(t=0)=\bfE_a}_{\bfE(t=T)=\bfE_b}}e^{\frac{i}{\hbar}S(T)},
\ee
where $S(T)$ is the action of \refeqn{Sem1} with the time integrals
taken between zero and $T$, and $\Ham$ is the corresponding Hamiltonian.

If the initial and final states are set equal and summed over, the
resulting functional integration defines the
Minkowski space functional integral
\be
\mathcal{Z}(T) \equiv \sum_{a}\bra{\bfE_a}e^{-i\Ham T/\hbar}\ket{\bfE_a}
=\tr e^{-i \Ham T/\hbar} = \int \dA \, 
e^{\frac{i}{\hbar} S(T)},
\labeleqn{tracetime}
\ee
which depends on the time $T$ and the background potential
$\tV(\omega,\vecx)$.  The partition function that describes this
system at temperature $1/\beta$ is defined by
\be
Z(\beta) = {\cal Z}(-i\hbar\beta) = \tr e^{-\beta \Ham},
\labeleqn{part}
\ee
and the free energy $F$ of the field is
\be
F(\beta) = -\frac{1}{\beta}\log Z(\beta).
\labeleqn{free}
\ee
The limit $\beta\to\infty$ projects the ground state energy out of the trace,
\be
 \calE_0 =  F(\beta = \infty) = 
-\lim_{\beta\to\infty} \frac{1}{\beta} \log  Z  {(\beta)}.
\labeleqn{E0}
\ee
The unrenormalized energy $\calE_0$ generally depends on an
ultraviolet cutoff, but  cutoff-dependent contributions arise from the
objects individually \cite{Graham03, Graham09} and do not depend on their separations or orientations.  
Such terms can remain after ordinary QED renormalization if objects are assumed
to constrain electromagnetic waves with arbitrarily high frequencies
(for example, if the fields are forced to vanish on a surface). Such
boundary conditions should be regarded as artificial idealizations;
in reality, when the wavelengths of the electromagnetic waves become
shorter than the length scales that characterize the interactions of
the material, the influence of the
material on the waves vanishes \cite{Graham03}. Accordingly, the potential 
$\tV$ should vanish for real materials
in the high-frequency limit. In any event these cutoff dependences
are independent of the separation and orientation of the objects, and since we are only interested in energy
\emph{differences}, we can remove them by subtracting the
ground state energy of the system when the objects are in some
reference configuration. In most cases we take this configuration
to be when the objects are infinitely far apart, but when calculating Casimir
energies for one object inside another, some other configuration must
be used.  We denote the partition function for this reference
configuration by $\overline Z$. In this way we obtain the Casimir energy,
\be
\calE = -\lim_{\beta\to\infty} \frac{1}{\beta} \log Z(\beta)/\overline
Z(\beta).
\ee
Throughout our calculation of $\calE$, we will thus be able to neglect
any overall factors that are independent of the relative positions and
orientations of the objects.

%%%%%%%%%%%%%%%%%%%%%%%%%%%%%%%%%%%%%%%%%%%%%%%%%%%%%%%%%%%%%%%%%%%
%%%%%%%%%%%%%%%%%%%%%%%%%%%%%%%%%%%%%%%%%%%%%%%%%%%%%%%%%%%%%%%%%%%
%\subsection{Euclidean Electromagnetic Action}
%\label{sec:EucEMaction}
 
By replacing the time $T$ by $-i\hbar\beta$, we transform the
Minkowski space functional integral $\mathcal{Z}(T)$ into the
partition function $Z(\beta)$.  In $A^0=0$ gauge, the result is
 simply to replace the frequencies
$\omega_n = \frac{2\pi n}{T}$ in \refeqn{Vem1} by $i\frac{2\pi n}{\hbar
\beta}=ic\kappa_n$, where $\kappa_n$ is
the $n^{\rm th}$  Matsubara frequency divided by $c$. (In other gauges
the temporal component $A^0$ of the vector field must be rotated too.)

The Lagrangian is quadratic, so the modes with different
$\kappa_n$ decouple and the partition function decomposes into a
product of partition functions for each mode.  Since 
the electromagnetic field is real, we have $\bfE^*(\omega) =
\bfE(-\omega)$ on the real axis.  We can thus further simplify this
decomposition on the imaginary axis by considering $\kappa\ge 0$ only,
but allowing $\bfE$ and $\bfE^*$ to vary independently in the path
integral.  Restricting to positive $\kappa$ is possible because the
response functions $\epsilon(ic\kappa,\vecx)$ and
$\mu(ic\kappa,\vecx)$ are invariant under a change of sign in
$ic\kappa$, as shown in Appendix~A of Ref.~\cite{Rahi09}. In the limit
$\beta\to\infty$, the sum $\sum_{n\geq 0}$ turns into an integral 
$\frac{\hbar c \beta}{2\pi}\int_{0}^\infty d\kappa$, and we have
\be
\calE_0 = -\frac{\hbar c}{2\pi} \int_0^\infty d\kappa \,
\log Z(\kappa),
\labeleqn{EKem}
\ee
where
\be
\begin{split}
Z(\kappa) = \int \dA \dA^* \, 
\exp & \left[ -\beta \int d\vecx \,
\bfE^{*} \cdot \left(\tI+\frac{1}{\kappa^2}
\curl \curl \right) \bfE +
\frac{1}{\kappa^2} \bfE^{*}  \cdot \tV(ic\kappa,\vecx) \, \bfE
 \right],
\end{split}
\labeleqn{ZKem}
\ee
\be
\tV(ic\kappa,\vecx) = \tI \, \kappa^2
\left(\epsilon(ic\kappa,\vecx)-1\right) + \curl
\left(\frac{1}{\mu(ic\kappa,\vecx)} -1 \right) \curl
\,.
\ee
The potential $\tV(ic\kappa,\vecx)$ is real for real $\kappa$, even
though $\epsilon$ and $\mu$ can have imaginary parts for real
frequencies $\omega$.  Our goal is now to manipulate $Z(\kappa)$ in
\refeqn{ZKem} so that it is computable from the scattering properties
of the objects.

%%%%%%%%%%%%%%%%%%%%%%%%%%%%%%%%%%%%%%%%%%%%%%%%%%%%%%%%%%%%%%%%%%%
%%%%%%%%%%%%%%%%%%%%%%%%%%%%%%%%%%%%%%%%%%%%%%%%%%%%%%%%%%%%%%%%%%%
%%%%%%%%%%%%%%%%%%%%%%%%%%%%%%%%%%%%%%%%%%%%%%%%%%%%%%%%%%%%%%%%%%%
\subsection{Green's function expansions and translation formulas}
\label{sec:Green}

The free Green's function and its representations in
various coordinate systems are crucial to our 
formalism.  The free electromagnetic field ($\tV=0$) obeys equations
of motion obtained by extremizing the corresponding action, \refeqn{Sem1},
\be
\left(- \tI \, \frac{\omega^2}{c^2} + 
\curl\curl \right)\bfE(\omega,\vecx) = 0.
\labeleqn{EoMem1}
\ee
We will employ the electromagnetic dyadic Green's function
$\tGzero$, defined by
\be
\left(- \tI \, \frac{\omega^2}{c^2}
+ \curl\curl \right)\tGzero(\omega,\vecx,\vecx') = 
\,\tI \delta^{(3)}\left(\vecx-\vecx'\right),
\labeleqn{gfeq}
\ee
written here
in the position space representation.
{The Green's
function has to be the retarded one, not only on physical grounds, but also as a consequence of the imaginary-frequency formalism, just as is the case for the response
functions $\epsilon$ and $\mu$. It is the \emph{retarded} response
functions that are analytically continued in the frequency domain to
positive imaginary frequency, as shown in Appendix~A of reference \cite{Rahi09}.}

{The representation of the free Green's function, which we need, employs the ``regular'' and ``outgoing''} solutions to the
differential equation, \refeqn{EoMem1},
\be
 \bfE^{\rm reg}_{\alpha}(\omega,\vecx) = \langle\vecx\ketEromaP,
\quad
 \bfE^{\rm out}_{\alpha}(\omega,\vecx) = \langle\vecx\ketEoutomaP,
\ee
represented formally by the eigenstate kets
$\ketEromaP$ and $\ketEromaP$, 
where the generalized index $\alpha$ labels the scattering channel,
including the polarization.  For example, for spherical wave functions
it represents the angular momentum quantum numbers $(l,m)$ and the
polarization $E$ or $M$. {There are six coordinate systems in which the vector wave equation (\ref{eq:EoMem1}) can be solved by separation of variables and vector wave functions appropriate to that coordinate system can be constructed~\cite{Morse53}. The labels ``regular'' and ``outgoing'' denote, respectively, the wave functions' non-singular behavior at the origin or `outward' direction of energy transport along one of the coordinate system's axes. Let us call the coordinate, along which the latter wave functions are outgoing $\xi_1$ and the other coordinates $\xi_2$ and $\xi_3$.}
We will usually work on the imaginary $\omega$-axis, in which case we will
encounter the corresponding modified special functions.

{The free Green's function can be expanded in tensor products of these wave functions,}
\be
\tGzero(\omega,\vecx,\vecx') =
\sum_{\aindex} C_{\aindex}(\omega)
\left\{
\begin{array}{l l}
\bfEoutaP(\omega,\xi_1,\xi_2,\xi_3)\otimes\bfEraPcc(\omega,\xi'_1,\xi'_2,\xi'_3)
 & \text{if } \xi_1(\vecx) > \xi'_1(\vecx') \\
\bfEraP(\omega,\xi_1,\xi_2,\xi_3)\otimes\bfEinaPcc(\omega,\xi'_1,\xi'_2,\xi'_3)
 & \text{if } \xi_1(\vecx) < \xi'_1(\vecx') \\
\end{array},
\right.
\labeleqn{G0expem}
\ee
$\bfEin_\alpha$ is the same as $\bfEoutaP$ except the functional dependence on $\xi_1$ is complex conjugated, making the wave function `incoming'. A list of
Green's function expansions in various common bases, including the normalization constant, $C_{\alpha}(\omega)$,
is given in Appendix~B of Ref.~\cite{Rahi09}.
The wave functions that appear in the series expansion of the free
Green's functions in \refeqn{G0expem} satisfy
wave equations with frequency $\omega$.
As we will see in 
Sec.~\ref{sec:Scatt}, the ability to express the Casimir energy
entirely in terms of an ``on-shell'' partial wave expansion with fixed
$\omega$ will greatly simplify our calculations. 

We will {also} use the free Green's function {in another representation }to combine the scattering amplitudes for two different
objects.  In this calculation  the one argument of the Green's
function will be located on each object. {As long as the pair of objects can be separated in one of the separable coordinate systems by the surface $\xi_1=\Xi=$const., we can distinguish an inside object which lies entirely inside the surface ($\xi_{1}<\Xi$) and an outside object ($\xi_{1}>\Xi$), see \reffig{transboth}.
\begin{figure}[ht]
\includegraphics[width=0.95\linewidth]{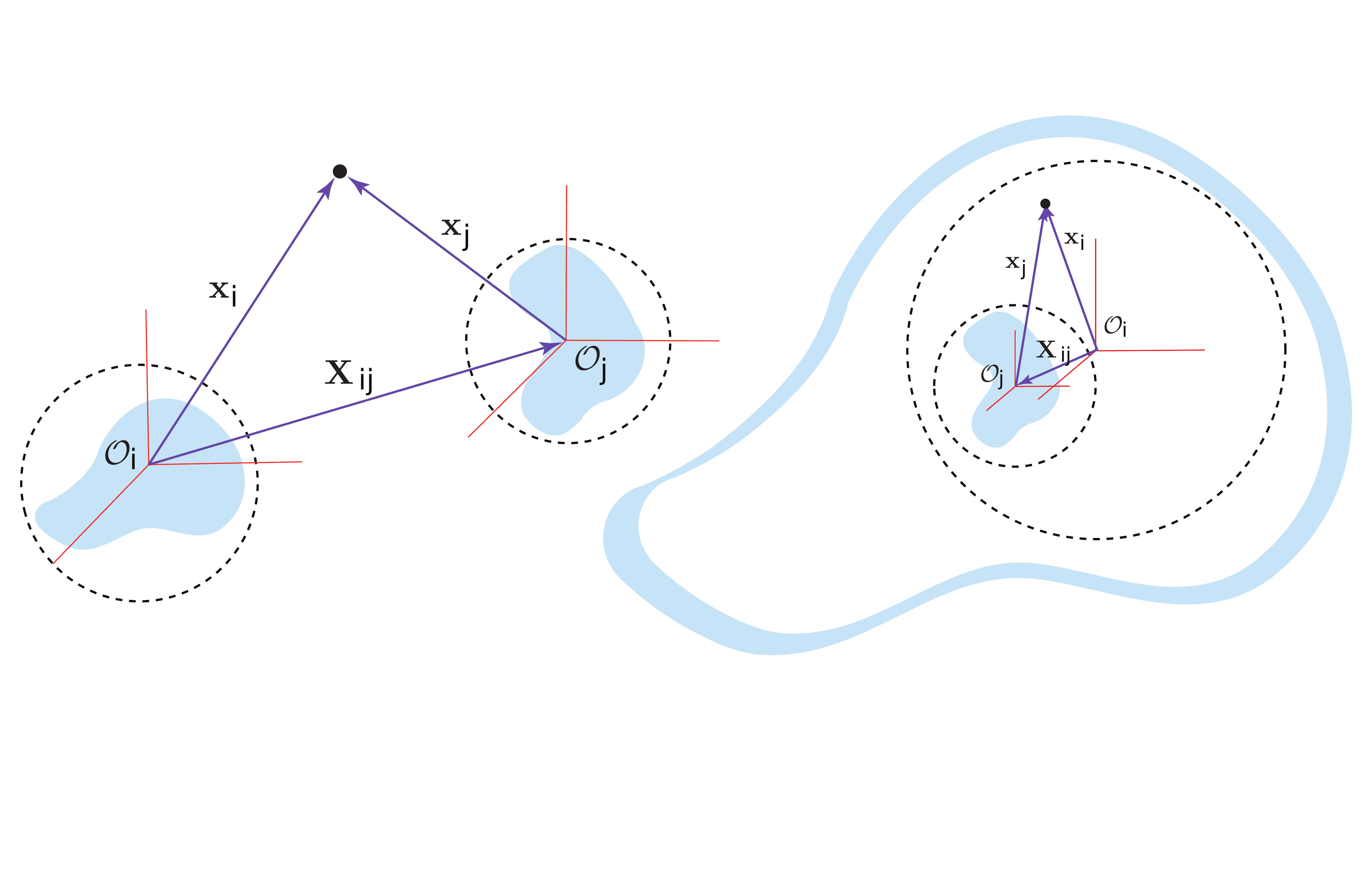}
\caption{
Geometry of the outside (left) and inside (right) configurations.
The dotted lines show surfaces separating the objects on which  the
radial variable is constant.  The translation vector
$\vecX_{ij} = \vecx_i - \vecx_j = -\vecX_{ji}$ describes the relative
positions of the two origins.
}
\label{fig:transboth}
\end{figure}
Then, we can expand the free Green's function, }when one argument, say
$\vecx$, lies on object $i$ and the other argument, say $\vecx'$,
lies on object $j$, we expand $\tGzero(ic\kappa,\vecx,\vecx')$ in
terms of coordinates $\vecx_i$ and $\vecx_j'$ that describe each point
relative to the origin of the body on which it lies.  {Which of the following expansions is appropriate for a particular pair of objects depends whether objects $i$ and $j$ are outside of one another, or one object is inside the other,}
\be
\begin{split}
& \tGzero(ic\kappa,\vecx,\vecx') =  \cr
& \sum_{\aindex,\bindex} C_\beta(\kappa)
\left\{
\begin{array}{l@{~}l}
\bfEregaP(\kappa,\vecx_i) \otimes \mathcal{U}^{ji}_{\abindex}(\kappa)
\bfEregbQcc(\kappa,\vecx_j')
&\hbox{if $i$ and $j$ are outside each other} \\
\bfEregaP(\kappa,\vecx_i) \otimes \mathcal{V}^{ij}_{\abindex}(\kappa)
\bfEinbQcc(\kappa,\vecx_j')&\left\{\begin{array}{l} 
 \hbox{if $i$ is inside $j$, or} \\ 
 \hbox{if $i$ is below $j$ (plane wave basis)} \end{array}\right.\\
\bfEoutaP(\kappa,\vecx_i) \otimes \mathcal{W}^{ji}_{\abindex}(\kappa)
\bfEregbQcc(\kappa,\vecx_j')&\left\{\begin{array}{l} 
 \hbox{if $j$ is inside $i$, or} \\
 \hbox{if $j$ is below $i$ (plane wave basis)}\end{array}\right. \\
\end{array} \right.
\end{split}
\labeleqn{G0cases}
\ee
where $\mathcal{W}^{ji}_{\alpha \beta}(\kappa) =
\mathcal{V}^{ji,\dagger}_{\alpha \beta} (\kappa)
\frac{C_{\aindex}(\kappa)}{C_{\bindex}(\kappa)}$ and $C_\aindex$ is
the normalization constant defined in \refeqn{G0expem}.
The expansion can be written more compactly as
\be
\tGzero(ic\kappa) = \sum_{\aindex,\bindex}(-C_{\bindex}(\kappa))
\left(\ketErkaaP ~~ \ketEoutkaaP\right)
\X^{ij}_{\abindex}(\kappa)
\left(
\begin{array}{c}
\braErkabQ \\
\braEinkabQ
\end{array}
\right),
\labeleqn{tG0shiftshort}
\ee
where  the $\X$ matrix is defined, for
convenience, as the negative of the matrix containing the translation
matrices,
\be
\begin{split}
\X^{ij}(\kappa) =
\left(
\begin{array}{c c}
-\mathcal{U}^{ji}(\kappa) & -\mathcal{V}^{ij}(\kappa) \\
-\mathcal{W}^{ji}(\kappa) & 0
\end{array}
\right).
\end{split}
\labeleqn{Xdef}
\ee
In \refeqn{tG0shiftshort} the bras and kets are to be evaluated in position space in the
appropriately restricted domains and only one of the three submatrices is nonzero for any
pair of objects $i$ and $j$ as given in \refeqn{G0cases}.
The translation matrices for various geometries are provided in
Appendix~C of reference \cite{Rahi09}.

%%%%%%%%%%%%%%%%%%%%%%%%%%%%%%%%%%%%%%%%%%%%%%%%%%%%%%%%%%%%%%%%%%%

%%%%%%%%%%%%%%%%%%%%%%%%%%%%%%%%%%%%%%%%%%%%%%%%%%%%%%%%%%%%%%%%%%%
%%%%%%%%%%%%%%%%%%%%%%%%%%%%%%%%%%%%%%%%%%%%%%%%%%%%%%%%%%%%%%%%%%%
\subsection{Classical scattering of electromagnetic fields}
\label{sec:Scatt}

In this section, we summarize the key results from scattering theory
needed to compute the scattering amplitude of each body individually.
In the subsequent section we
will then combine these results with the translation matrices of the
previous section to compute $Z(\kappa)$.

{By combining the frequency-dependent Maxwell equations, one obtains the vector wave equation}
\be
(\Hzero + \V(\omega,\vecx))\bfE(\omega,\vecx) = \frac{\omega^2}{c^2} 
\bfE(\omega,\vecx),
\labeleqn{EoMem2}
\ee
where
\be
\begin{gathered}
\labeleqn{gathered1}
\Hzero = \curl \curl , \\
\V(\omega,\vecx) = \tI \frac{\omega^2}{c^2} \minepnew + \curl \minmunew \curl,
\end{gathered}
\ee
which is the same potential operator as the one obtained by rearranging the
Lagrangian (see \refeqn{Vem1}).

The Lippmann-Schwinger equation \cite{Lippmann50}
\be
\ketE = \ketEh - \tGzero \V \ketE
\labeleqn{LS}
\ee
expresses the general solution to \refeqn{EoMem2}.  Here{,}
$\tGzero$ is the free electromagnetic tensor Green's function
discussed in Sec. \ref{sec:Green} and the homogeneous solution $\ketEh$
obeys $\left(-\frac{\omega^2}{c^2}\tI + \Hzero\right) \ketEh = 0$.  We can iteratively substitute for
$\ketE$ in \refeqn{LS} to obtain the
formal expansion
\be
\begin{split}
\ketE &  = \ketEh - \tGzero \V \ketEh + \tGzero \V \tGzero \V \ketE -
\ldots \\
& = \ketEh - \tGzero \T \ketEh ,
\end{split}
\labeleqn{LSEM}
\ee
where the electromagnetic $\T$-operator is defined as
\be
\T = \V \frac{\tI}{\tI + \tGzero \V} = \V \tG \tGzero^{-1},
\labeleqn{Tem}
\ee
and $\tG$ is the Green's function of the full Hamiltonian,
$\left(-\frac{\omega^2}{c^2} \tI + \Hzero + \V
\right)\tG=\tI$.  We note that  $\T$, $\tGzero$, and $\tG$ are all
functions of frequency $\omega$ and non-local in
space.  As can be seen from expanding $\T$ in \refeqn{Tem} in a power
series, $\T(\omega,\vecx,\vecx')=\bra{\vecx} \T(\omega) \ket{\vecx'}$ is
zero whenever $\vecx$ or $\vecx'$ are not located on an object, {\it
i.e.\/}, where $\V(\omega,\vecx)$ is zero.  This result does not, however,
apply to
\be
\T^{-1} = \tGzero + \V^{-1},
\labeleqn{tinverse}  
\ee
because the free Green's function is nonlocal.  The potential $\V(\omega,\vecx)$ which appears in \refeqn{gathered1} is the coordinate space matrix element of $\V$, $\bra{\vecx}\V\ket{\vecx'}=\V(\omega,\vecx)\delta(\vecx-\vecx')$, which can be generalized to the case where $\V$ is non-local, $\bra{\vecx}\V\ket{\vecx'}=\V(\omega,\vecx,\vecx')$.  Note that whether $\V$ is local or non-local, its matrix elements vanish if $\vecx$ and $\vecx'$ are on different objects or if either $\vecx$ or $\vecx'$ is outside of the objects.  The definition of $\V^{-1}$ is natural, $\bra{\vecx}\V^{-1}\ket{\vecx'}=\V^{-1}(\omega,\vecx)\delta(\vecx-\vecx')$ (and similarly for the non-local case) when $\vecx$ and $\vecx'$ are on a single object, which is the only case that enters our analysis.
 
Next we connect the matrix elements of the $\T$-operator between
states with equal $\omega$ to the scattering amplitude $\f$.  In our
formalism, only this restricted subset of $\T$-operator matrix
elements is needed in the computation of the Casimir energy.

\begin{figure}[htb]
\hfill
\includegraphics[width=0.4\linewidth]{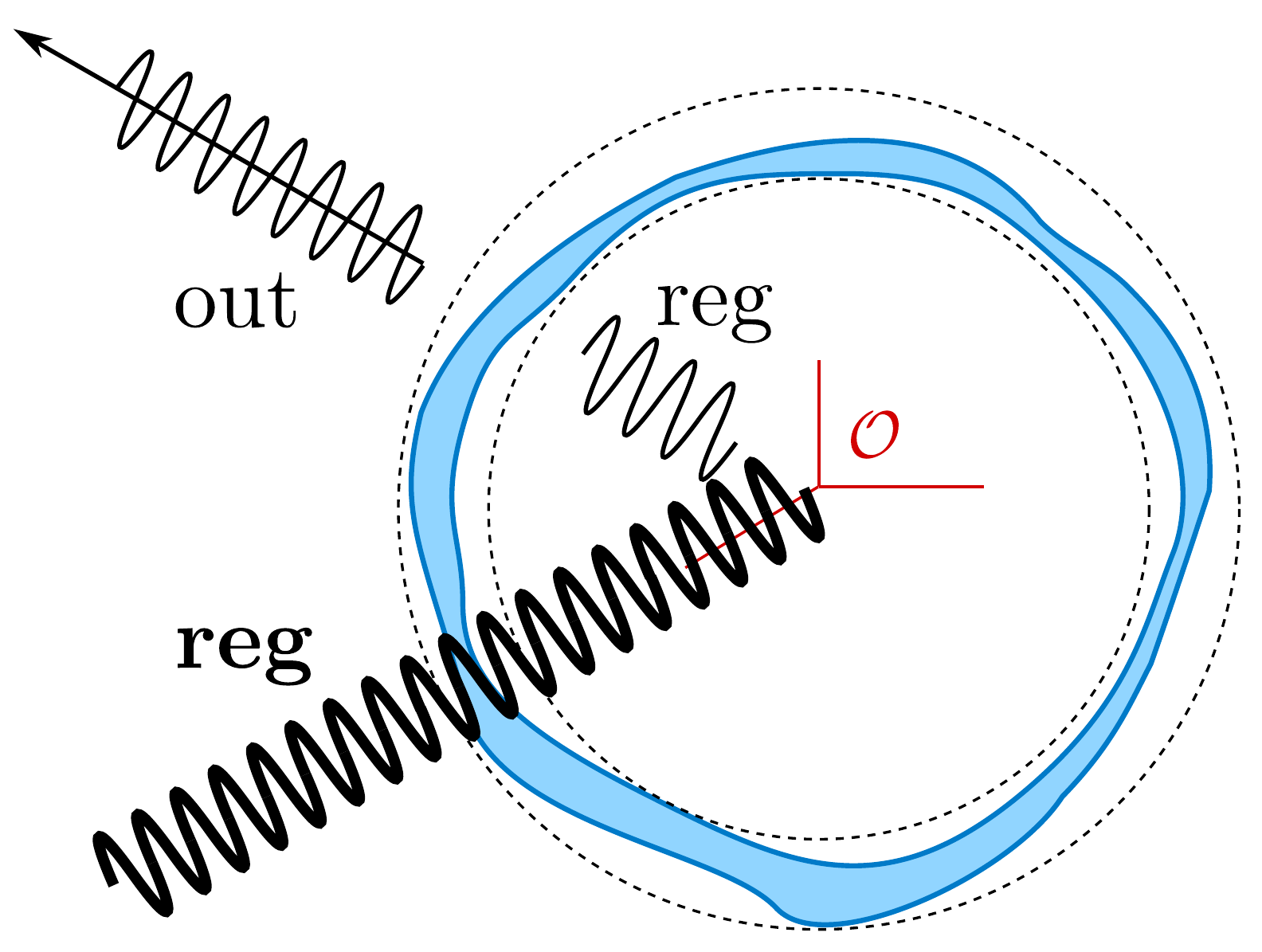}
\hfill
\includegraphics[width=0.4\linewidth]{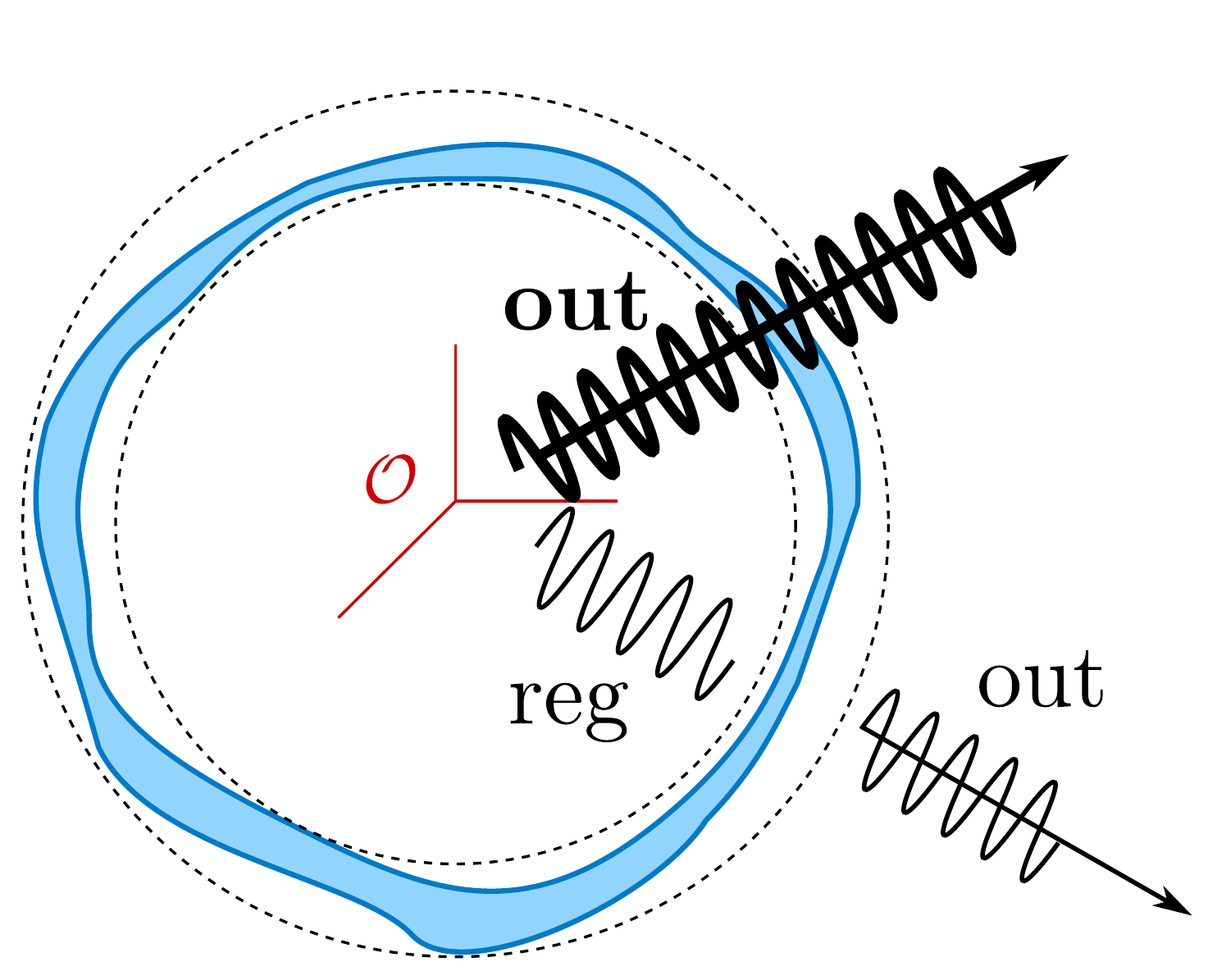}
\hfill
\caption{
The scattering waves for outside scattering (left panel) and inside
scattering (right panel).  In both cases
the homogeneous solution $\bfEhom$ is shown in bold.  For outside
scattering, the homogeneous solution is a regular wave,
which produces a regular wave inside the object and an outgoing wave
outside the object.  For inside scattering, the
homogeneous solution is an outgoing wave, which produces a regular
wave inside the object and an outgoing wave outside the object.
}
\label{fig:scatt}
\end{figure}

By the choice of the homogeneous solution, $\ketEh$, is regular or outgoing, we can distinguish two physically different processes. In the former case, the object scatters the regular wave outward and modifies the amplitude of the imposed regular wave functions inside, a situation we refer to as outside scattering (left panel of \reffig{scatt}).\footnote{
Alternatively, we can set up asymptotically incoming and outgoing
waves on the outside and regular waves inside. The amplitudes of the
outgoing waves are then given by the $ S$-matrix, which is related to
the scattering amplitude $\f$ by $\f=( S- I)/2$.  Although these two
matrices carry equivalent information, the scattering amplitude will
be more convenient for our calculation.} In the latter case, the object modifies the amplitude of the transmitted wave and partly reflects it as a regular wave inside (inside scattering, right panel of \reffig{scatt}). `Outside' and `inside' are distinguished by surfaces $\xi_1$=constant, as before. Here we treat the \emph{outside scattering case}, and refer the reader to Ref.~\cite{Rahi09} the \emph{inside case} and further details. The expansion in \refeqn{G0expem} allows us to express \refeqn{LSEM} as
%\be
%\bfE(\omega, \vecx)  =  \bfEregaP(\omega, \vecx) - 
%\sum_{\bindex} \bfEoutbQ(\omega, \vecx) \int
%C_{\bindex}(\omega) \bfEregbQcc(\omega, \vecx') \cdot
%\T(\omega, \vecx',\vecx'') \bfEregaP(\omega, \vecx'') d\vecx'd\vecx''.
%\ee
\begin{align}
\bfE(\omega, \vecx)  =  \bfEregaP(\omega, \vecx) & - 
\sum_{\bindex} \bfEoutbQ(\omega, \vecx) \\
& \times \int
C_{\bindex}(\omega) \bfEregbQcc(\omega, \vecx') \cdot
\T(\omega, \vecx',\vecx'') \bfEregaP(\omega, \vecx'') d\vecx'd\vecx''. \nonumber
\end{align}
{at points $\vecx$ outside a surface $\xi_1=$constant enclosing the object.} 
The equation can be written in Dirac notation, again with the
condition that the domain of the functional Hilbert space 
is chosen appropriately to the type of solution, 
\be
\ketEom = \ketEromaP + \sum_{\bindex} \ketEoutombQ \times
\underbrace{(-1) C_{\bindex}(\omega) \braErombQ \T (\omega)\ketEromaP}_{
\Tregreg_{\bindex, \aindex}(\omega)},
\labeleqn{Temii}
\ee
which defines $\Tregreg_{\beta,\alpha}$ as the exterior/exterior
scattering amplitude (the one evaluated between two
regular solutions).  We will use analogous notation in the other
cases below. 

At coordinates $\vecx$ ``far enough inside'' a cavity of the object,
meaning that $\vecx$ has smaller $\xi_1$ than any point on the object,
the field $\bfE$ is given by
\be
\ketEom = \ketEromaP + \sum_{\bindex} \ketErombQ \times
\underbrace{(-1) C_{\bindex}(\omega) \braEinombQ
\T (\omega)\ketEromaP}_{
\Toutreg_{\bindex, \aindex}(\omega)},
\labeleqn{Temoi}
\ee
where again the free states are only defined over the appropriate
domain in position space, and $\Toutreg$ indicates the
interior/exterior scattering amplitude.

We have obtained the scattering amplitude in the basis of free
solutions with fixed $\omega$.  Since one is normally interested in
the scattering of waves outside the object, the scattering amplitude
usually refers to  $\Tregreg$.  We will use a more
general definition, which encompasses all possible combinations of
inside and outside.  The scattering amplitude is always ``on-shell,''
because the frequencies of both the operator and the states is $\omega$.
As a result, it is a special case of the
$\T$-operator, which {can connect} wave functions with different $\omega$.

We find it convenient to assemble the scattering amplitudes for
inside and outside into a single matrix,
\be
\begin{split}
\F(\kappa) &=
\left(
\begin{array}{c c}
\Tregreg(\kappa) & \Tregout(\kappa) \\
\Toutreg(\kappa) & \Toutout(\kappa)
\end{array}
\right) \\
& =  (-1)C_{\aindex}(\kappa)
\left(
\begin{array}{c c}
\braErkaaP \T(ic\kappa) \ketErkabQ & \braErkaaP \T(ic\kappa)
\ketEoutkabQ \\ \braEinkaaP \T(ic\kappa) \ketErkabQ & \braEinkaaP
\T(ic\kappa) \ketEoutkabQ
\end{array}
\right).
\end{split}
\labeleqn{Temdef}
\ee
where we have set $\omega=ic\kappa$, since this is the case we use.
For simplificity we define  $\left.\Toutreg_{\bindex, \aindex}(\omega)\right|_{\omega=ic\kappa}\equiv \Tregreg_{\bindex,\aindex}(\kappa)$.

%%%%%%%%%%%%%%%%%%%%%%%%%%%%%%%%%%%%%%%%%%%%%%%%%%%%%%%%%%%%%%%%%%%
%%%%%%%%%%%%%%%%%%%%%%%%%%%%%%%%%%%%%%%%%%%%%%%%%%%%%%%%%%%%%%%%%%%
\subsection{Casimir free energy in terms of the scattering amplitudes}
\label{sec:parttoscat}
\noindent

With the tools of the previous two sections, we are now able to 
re-express the Euclidean electromagnetic partition function of
\refeqn{ZKem} in terms of the scattering theory results derived in
Section \ref{sec:Scatt} for imaginary frequency.
We   exchange the fluctuating field $\A$, which
is subject to the potential $\V(ic\kappa,\vecx)$, for 
a free field $\A'$, together with fluctuating currents $\J$ and
charges $-\frac{i}{\omega}\nabla\cdot\J$ that are confined to
the objects\footnote{The sequence of two changes of
variables is known as
Hubbard-Stratonovich transformation in condensed matter physics.}

We multiply and divide the partition function \refeqn{ZKem} by
\be
\begin{split}
W & = \int \dJJ \exp\left[-\beta
\int d\vecx \, \J^*(\vecx) \cdot \V^{-1}(ic\kappa,\vecx)
\J(\vecx) \right]  = \det \V(ic\kappa,\vecx,\vecx')\, ,
\end{split}
\ee
where $\left.\right|_{\rm obj}$ indicates
that the currents are defined only over the objects, {\it i.e.\/} the
domain where $\V$ is nonzero  and therefore $\V^{-1}$ exists.

We then change variables in the integration,
$\J(\vecx) = \J'(\vecx) + \frac{i}{\kappa} 
\V(ic\kappa,\vecx) \bfE(\vecx)${, and a second time,}
$\bfE(ic\kappa,\vecx) = \bfE'(ic\kappa,\vecx) - i \kappa \int
d\vecx' \, \tGzero(ic\kappa,\vecx,\vecx') \J'(\vecx')$ and analogously for $\J^{*}$ and $\bfE^{*}$, to obtain
\be
\begin{split}
Z(\kappa) & = \frac{Z_0}{W} \int
\dJJprime \cr
& \exp\left[
-\beta \int d\vecx d\vecx' \,
{\J'}^{*}(\vecx) \cdot \left(
\tGzero(ic\kappa,\vecx,\vecx')
 + \V^{-1}(ic\kappa,\vecx, \vecx')\right) \J'(\vecx')
\right],
\end{split}
\labeleqn{Zemfull}
\ee
where
\be
Z_0 = \int \dA' \dA'^*
\exp\left[{-\beta \int d\vecx \,
{\bfE'}^*(\vecx)\cdot
\left(\tI +  \frac{1}{\kappa^2}\curl\curl \right)\bfE'(\vecx)} \right]
\ee
is the partition function of the free field, which is independent of
the objects. In $Z(\kappa)$, current fluctuations replace the field
fluctuations of \refeqn{ZKem}. The interaction of current fluctuations on different objects
is described by the free Green's function
$\tGzero(ic\kappa,\vecx,\vecx')$ alone. The inverse potential penalizes
current fluctuations if the potential is small.

To put the partition function into a suitable form for practical
computations, we use the results of the previous sections to 
re-express the microscopic current fluctuations
as macroscopic multipole fluctuations,  which then can be
connected  to the individual objects' scattering amplitudes. 
This transformation comes about naturally once
the current fluctuations are decomposed according to the objects on
which they occur and the appropriate expansions of the
Green's function are introduced.  We begin this process by
noticing that the operator in the exponent of the integrand in
\refeqn{Zemfull} is the negative of the inverse of the $\T$-operator
(see \refeqn{tinverse}), and hence
\be
Z(\kappa) =
Z_0 \, 
\det \V^{-1}(ic\kappa,\vecx,\vecx') \, 
\det \T(ic\kappa,\vecx,\vecx')
\labeleqn{ZemT}
\ee
which is in agreement with a more formal calculation:  Since
 $Z_0 = \det \tG_0(ic\kappa,\vecx,\vecx')$ and
$Z(\kappa) = \det \tG(ic\kappa,\vecx,\vecx')$,
 we only need to take the determinant of \refeqn{Tem} to arrive at the
result of \refeqn{ZemT}.

Both $Z_0$ and $\det \V^{-1}(ic\kappa,\vecx)$ are independent of the
separation of the objects, since the former is simply the free Green's
function, while the latter is diagonal in $\vecx$.  Even a nonlocal
potential $\V(ic\kappa,\vecx,\vecx')$ only connects points within the same
object, so its determinant is also independent of the objects'
separation.  Because these determinants do not depend on
separation, they are canceled by a reference partition function in
the final result.  We are thus left with the task of computing the
determinant of the $\T$-operator.

As has been discussed in Sec. \ref{sec:Scatt}, the $\T$-operator
$\T(ic\kappa,\vecx,\vecx')$ is not diagonal in the spatial
coordinates. Its determinant needs to be taken over the spatial
indices $\vecx$ and $\vecx'$, which are restricted to the objects
because the fluctuating currents $\J(\vecx)$ in the functional
integrals are zero away from the objects.  This determinant also runs
over the ordinary vector components of the electromagnetic $\T$ operator.

A change of basis to momentum space does not help in computing
the determinant of the $\T$-operator, even though it does
help in finding the determinant of the free Green's function{, for example}. One
reason is that the momentum basis is not orthogonal over the domain of
the indices $\vecx$ and $\vecx'$, which is restricted to the objects. In
addition, a complete momentum basis includes not only all directions
of the momentum vector, but also all magnitudes of the momenta. So, in
the matrix element $\bra{\bfE_\veck} \T(\omega) \ket{\bfE_{\veck'}}$ the
wave numbers $k$ and $k'$ would not have to match, and could also
differ from $\omega/c$.  That is, the matrix elements could be
``off-shell.''  Therefore, the $\T$-operator could not simply be treated
as if it was the scattering amplitude, which is the on-shell representation of
the operator in the subbasis of frequency $\omega$
(see Sec. \ref{sec:Scatt}), and is significantly easier to calculate.
Nonetheless, we will see that it is possible to express the Casimir
energy in terms of the on-shell operator only, by remaining in the
position basis.

From \refeqn{Tem}, we know that the inverse of the $\T$-operator
equals the sum of the free Green's function and the inverse of the
potential.  Since the determinant of the inverse operator is the
reciprocal of the determinant, it is expedient to start with the
inverse $\T$-operator.  We then separate the basis involving all the
objects into blocks for the $n$ objects.  In a schematic notation, we have
\be
[\brax \T^{-1} \ketxp] = \left(
\begin{array}{c|c|c}
[\braxone \T_1^{-1} \ketxonep] & 
[\braxone \tGzero \ketxtwop] & \cdots \\ \hline
[\braxtwo \tGzero \ketxonep] & 
[\braxtwo \T_2^{-1} \ketxtwop] & \cdots \\ \hline
\cdots & \cdots & \cdots
\end{array}
\right),
\ee
where the $ij^{\rm th}$ submatrix refers to $\vecx\in$ object $i$ and
$\vecx'\in$ object $j$ and $\vecx_i$ represents a point in object $i$
measured with respect to some fixed coordinate system. Unlike the position
vectors in Sec. \ref{sec:Green}, at this point the subscript of
$\vecx_i$ does not indicate the origin with respect to which the
vector is measured, but rather the object on which the point
lies. Square brackets are used to remind us that we are considering
the entire matrix or submatrix and not a single matrix element.  We
note that the operators $\T$ and $\tGzero$ are functions of
$ic\kappa$, but for simplicity we suppress this argument throughout
this derivation.  When the two spatial indices lie on different
objects, only the free Green's function remains in the off-diagonal
submatrices, because   $\langle  \vecx_{i}|\V^{-1}|
\vecx'_{j}\rangle=0$ for $i\ne j$.

Next, we multiply $\T^{-1}$ by a reference $\T$-operator $\T_\infty$
without off-diagonal submatrices, which can be interpreted as the
$\T$-operator at infinite separation,
\be
\begin{split}
& [\brax \T_\infty  \T^{-1} \ketxpp] = \\
& \left(
\begin{array}{c|c|c}
[\langle \vecx_1 \ketxonepp] &
[\int d\vecx_1' \, \braxone \T_1 \ketxonep \braxonep \tGzero \ketxtwopp] &
\cdots \\ \hline
[\int d\vecx_2' \, \braxtwo \T_2 \ketxtwop \braxtwop \tGzero \ketxonepp] &
[\langle \vecx_2 \ketxtwopp] &
\cdots \\ \hline
\cdots & \cdots & \cdots
\end{array}
\right).
\end{split}
\labeleqn{TTinv}
\ee
Each off-diagonal submatrix $[\int d\vecx_i' \braxi \T_i \ketxip \braxip
\tGzero \ketxjpp]$ is the product of the $\T$-operator of object $i$,
evaluated at two points $\vecx_i$ and $\vecx_i'$ on that object,
multiplied by the free Green's function, which connects $\vecx_i'$
to some point $\vecx_j''$ on object $j$.

Now we shift all variables to the coordinate systems of the objects on
which they lie.  As a result, the index on a position vector
$\vecx_i$ now refers to the object $i$ on which the point lies 
\emph{and} to the coordinate system with origin $\orig_i$ in which the
vector is represented, in agreement with the notation of 
Sec. \ref{sec:Green}.  The off-diagonal submatrices in
\refeqn{TTinv} can then be rewritten using \refeqn{tG0shiftshort} as,
\be
\sum_{\aindex, \bindex}
\left[
\left(\braxi \T_i\ketErkaaP ~~ \braxi \T_i\ketEoutkaaP \right)
\X^{ij}_{\abindex}
\left(
\begin{array}{c}
\braErkabQ \vecx_j'' \rangle \\
\braEinkabQ \vecx_j'' \rangle
\end{array}
\right) (-C_{\bindex}(\kappa))
\right].
\ee

The matrix $[\brax \T_\infty \T^{-1} \ketxpp]$ has the structure 
$\tI + \tA\tB${. }Using Sylvester's determinant formula
$\det(\tI+\tA\tB)=\det(\tI+\tB\tA)$, we see that 
the determinant is unchanged if we
replace the off-diagonal submatrices in \refeqn{TTinv} by
\be
\left[
\sum_{\bindex}
(-1)C_{\aindex}(\kappa)
\left(
\begin{array}{c c}
\braErkaaP \T_i \ketErkabQ & 
\braErkaaP \T_i \ketEoutkabQ \\
\braEinkaaP \T_i \ketErkabQ & 
\braEinkaaP \T_i \ketEoutkabQ
\end{array}
\right)
\X^{ij}_{\bindex, \cindex}
\right].
\labeleqn{TW}
\ee
With this change, the diagonal submatrices in \refeqn{TTinv} become
diagonal in the partial wave indices rather than in position
space. The matrix elements of the $\T$-operator are the scattering
amplitudes, which can be obtained from ordinary scattering calculations,
as demonstrated in Sec. \ref{sec:Scatt}.   The first matrix in
\refeqn{TW}, including the prefactor $(-1)C_{\aindex}(\kappa)$, is
$\F_i(\kappa)$, the modified scattering amplitude of object $i$, defined in
\refeqn{Temdef}.

Putting together Eqs. \refeq{EKem}, \refeq{ZKem}, \refeq{ZemT}, and
\refeq{TTinv}, we obtain
\begin{equation}
\label{Elogdet}
\mathcal{E} = \frac{\hbar c}{2\pi} \int_0^\infty d\kappa 
\log \det (\mathbb{M} \mathbb{M}_\infty^{-1}),
\end{equation}
where
\be
\mathbb{M} =
\left(
\begin{array}{c c c c}
\F_1^{-1} & \X^{12} & \X^{13} & \cdots \\
\X^{21}    & \F_2^{-1} & \X^{23} & \cdots \\
\cdots & \cdots & \cdots & \cdots
\end{array}
\right)
\ee
and $\mathbb{M}^{-1}_\infty$ is a block diagonal matrix
$\text{diag}(\F_1 ~~\F_2 ~\cdots)$.

Using the block determinant identity 
\be
\det
\left(
\begin{array}{c c}
\tA & \tB \\
\tC & \tD
\end{array}
\right)
=
\det \left(\tA\right)
\det \left(\tD-\tC \tA^{-1} \tB\right)
=
\det \left(\tD\right)
\det \left(\tA-\tB \tD^{-1} \tC \right),
\labeleqn{detblock}
\ee
we can simplify this expression for the case of the interaction
between two objects,
\be
\mathcal{E} = \frac{\hbar c}{2\pi} \int_0^\infty d\kappa \log \det
\left(\tI - \F_a\X^{ab}\F_b\X^{ba}\right).
\labeleqn{Elogdet2}
\ee

Usually, not all of the submatrices of $\F$ and $\X$ are actually
needed for a computation. For example, if all objects are outside of
one another, only the submatrices $\Tregreg$ of the scattering amplitude
that describe outside reflection are needed.  If there are only two
objects, one inside another, then only the inside reflection submatrix
$\Toutout$ of the outside object and the outside reflection submatrix
$\Tregreg$ of the inside object are needed.

In order to obtain the free energy at nonzero temperature instead of
the ground state energy, we do not take the limit $\beta\to\infty$ in
\refeqn{E0}.  Instead, the integral $\frac{\hbar
c}{2\pi} \int_0^\infty d\kappa$ is replaced everywhere by $\frac{1}{\beta}
\sum_{n}'$, where $c \kappa_n=\frac{2\pi n}{\hbar\beta}$ with
$n=0,1,2,3\ldots$ is the $n$th Matsubara frequency.  A careful
analysis of the derivation shows that the zero frequency mode is
weighted by $1/2$ compared to the rest of the terms in the sum; this
modification of the sum is denoted by a prime on the summation
symbol. The factor of $1/2$ comes about because the fluctuating
charges or currents have to be real for zero frequency. Thus, for
$\kappa_0$, the expressions on the right hand side of
\refeqn{ZemT} should be placed under a square
root.  (For a complex field, both signs of the integer $n$ would be
included separately, and $n=0$ would be included once, with the normal
weight.)

If the medium between the objects is not vacuum but instead has
permittivity $\epsilon_M(ic\kappa)$ and magnetic
permeability $\mu_M(ic\kappa)$ different from unity, then the free
Green's function is multiplied by $\mu_M(ic\kappa)$, and its argument
$\kappa$ is replaced by $n_M(ic\kappa)\kappa$, where
$n_M(ic\kappa)=\sqrt{\epsilon_M(ic\kappa)\mu_M(ic\kappa)}$ is the
medium's index of refraction.  Effectively, this change just scales
all frequency dependencies in the translation matrices $\X(\kappa)$,
which become $\X\left(n_M(ic\kappa)\kappa\right)$.  Furthermore, the
scattering amplitudes absorb the factor $\mu_M(ic\kappa)$ from the
free Green's function and change non-trivially, {\it i.e.\/}
not just by some overall factor or a scaling of the frequency.  They
have to be computed with the nonzero electric and magnetic
susceptibilities of the medium.

%% file: chap3.tex
%%%%%%%%%%%%%%%%%%%%%%%%%%%%%%%%%%%%%%%%%%%%%%%%%%%%%%%%%%%%%%%%%%%
%%%%%%%%%%%%%%%%%%%%%%%%%%%%%%%%%%%%%%%%%%%%%%%%%%%%%%%%%%%%%%%%%%%
\section{Constraints on stable equilibria}
\label{sec:Constraints}

{Before presenting particular applications of the Casimir energy expression in Eq.~(\ref{Elogdet}), we consider some general properties of electrodynamic Casimir interactions here. This section has been adapted from a letter, which is co-authored by two of us~\cite{Rahi09-3}.}

\begin{figure}[htbp]
\begin{center}
\includegraphics[width=0.4\linewidth]{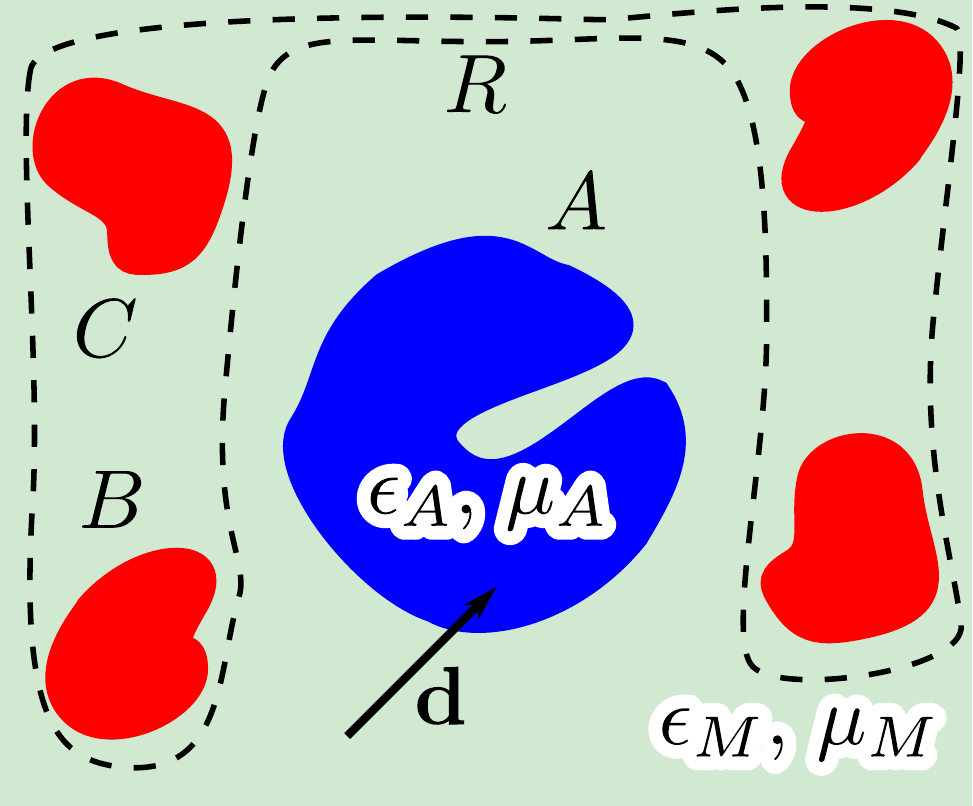}
\caption{The Casimir energy is considered for objects with electric permittivity $\epsilon_i(\omega,\vecx)$ and magnetic permeability $\mu_i(\omega,\vecx)$, embedded in a 
medium with uniform, isotropic, $\epsilon_M(\omega)$ and $\mu_M(\omega)$. 
To study the stability of object $A$, the rest of the objects are grouped in the combined entity $R$. The stability of the position of object $A$ is probed by displacing it infinitesimally by vector $\vecd$.}
\label{fig:all}
\end{center}
\end{figure}

As described in the Introduction, some general statements about the attractive or repulsive nature of Casimir forces can be made on the basis of the relative permittivity and permeability of objects and the medium they are immersed in.
But the sign of the force is largely a matter of perspective, since attractive forces can be easily arranged to produce repulsion along a specific direction, e.g., as in Ref.~\cite{rodriguez-2008}.  Instead, we focus on the question of stability{, see \reffig{all}}, which is more relevant to the design and development of MEMs and levitating devices.  We find that interactions between objects within the same class of material (as defined in the Introduction) cannot produce stable configurations.

{Let us take a step back and consider the question of stability of mechanical equilibria in the realm of electromagnetism.} 
Earnshaw's theorem~\cite{Earnshaw42} states that a collection of charges 
cannot be held in stable equilibrium solely by electrostatic forces. 
The charges can attract or repel, but cannot be stably levitated. 
While the stability of matter (due to quantum phenomena) {is a} vivid reminder of the caveats to this theorem,
it remains a powerful indicator of the constraints to stability in electrostatics.
An extension of Earnshaw's theorem to polarizable objects by Braunbek~\cite{Braunbek39-1,Braunbek39-2} establishes that dielectric and paramagnetic ($\ep>1$ and $\mu>1$) matter cannot be stably levitated by electrostatic forces, while diamagnetic ($\mu<1$) matter can. 
This is impressively demonstrated by superconductors and frogs that fly freely above magnets~\cite{Geim98}. 
If the enveloping medium is not vacuum, the criteria for stability are modified by substituting the static electric permittivity $\ep_M$ and magnetic permeability $\mu_M$ of the medium in place of the vacuum value of $1$ in the respective inequalities. 
In fact, if the medium itself has a dielectric constant higher than the objects ($\ep<\ep_M$), stable levitation  is possible, as demonstrated  for bubbles in liquids (see Ref.~\cite{Jones95}, and references therein).
For dynamic fields the restrictions of electrostatics do not apply; for example, lasers can lift and hold dielectric beads with index of refraction $n=\sqrt{\ep\mu}>1$~\cite{Ashkin70}.
{In addition to the force which keeps the bead in the center of the laser beam there is radiation pressure which pushes the bead along the direction of the Poynting vector. 
Ashkin and Gordon have proved that no arrangement of lasers can stably levitate an object just based on radiation pressure~\cite{Ashkin83}.
}

We begin our analysis of equilibria of the electrodynamic Casimir force with the precursor of Eq.~(\ref{Elogdet}), which contains the abstract $\T$ and $\tGM$-operators, where $\tGM $ is the electromagnetic Green's function operator for an isotropic, homogeneous medium,~\footnote{$\tGM$ satisfies $\left(\curl\mu_M^{-1}(ic\kappa)\curl+\ep_M(ic\kappa)\kappa^2\right)\tGM(ic\kappa,\vecx,\vecx')=\delta(\vecx-\vecx')\tI$, and is related to $G_M$, the Green's function of the imaginary frequency Helmholtz equation, by $\tGM(ic\ka,\vecx,\vecx') = \mu_M(ic\kappa)\left(\tI + (n_M \ka)^{-2} \boldsymbol{\nabla}\otimes\boldsymbol{\nabla}'\right) G_M(icn_M\ka,\vecx,\vecx')$.~Here, $n_M(ic\ka)=\sqrt{\ep_M(ic\ka) \mu_M(ic\ka)}$ is the index of refraction of the medium, whose argument is suppressed to simplify the presentation. {Thus $\tGM$, in contrast to $\tGzero$, takes into account the permittivity and permeability of the medium when they are different from one.}} 
\be
\mathcal{E} = \frac{\hbar c}{2\pi} \int_0^\infty d\ka\,\,\tr\ln \T^{-1} \T_\infty 
\, ,
\labeleqn{ECasimir}
\ee
where the operator $[\T^{-1}(ic\kappa,\vecx,\vecx')]$ equals
\be
\left(
\begin{array}{ c c c }
\left[\T_A^{-1}(ic\kappa,\vecx_1,\vecx'_1)\right] & \left[\tGM     (ic\kappa,\vecx_1,\vecx'_2)\right] & \cdots \\
\left[\tGM     (ic\kappa,\vecx_2,\vecx'_1)\right] & \left[\T_B^{-1}(ic\kappa,\vecx_2,\vecx'_2)\right] &  \\
\cdots   & & \cdots \\
\end{array}
\right)\,,
\ee
and $\T_\infty$ is the inverse of $\T^{-1}$ with $\tGM$ set to zero. The square brackets ``[ ]'' denote the entire matrix or submatrix with rows indicated by $\vecx$ and columns by $\vecx'$.
\footnote{To obtain the free energy at finite temperature, in place of the ground state energy $\mathcal{E}$,
 $\int \frac{d\ka}{2\pi}$ is replaced by the sum $\frac{kT}{\hbar c}\sum'_{\ka_n\geq 0}$ over Matsubara `wavenumbers' $\ka_n = 2\pi n k T/\hbar c$
 with the $\kappa_0=0$ mode weighted by $1/2$.}
 The operator $[\T^{-1}(ic\kappa,\vecx,\vecx')]$ has indices in position space. Each spatial index is limited to lie inside the objects $A,B,\cdots$. 
 For both indices $\vecx$ and $\vecx'$ in the same object $A$ the operator is just the inverse $\T$ operator of that object, $[\T_A^{-1}(ic\kappa,\vecx,\vecx')]$. 
For indices on different objects, $\vecx$ in $A$ and $\vecx'$ in $B$, it equals the electromagnetic Green's function operator $[\tGM(ic\kappa,\vecx,\vecx')]$ for an isotropic, homogeneous medium. 

As shown {in section \ref{sec:parttoscat}}, after a few manipulations, the operators $\T_J$ and $\tGM$ turn into the on-shell scattering amplitude matrix, $\mathbb{F}_J$, of object $J$ and the translation matrix $\X$, which converts wave functions between the origins of different objects. 
While practical computations require evaluation of the matrices in a particular {wave function} basis, the {position space} operators $\T_J$ and $\tGM$ are better suited to our general discussion here.

To investigate the stability of object $A$, we group the `rest' of the objects into a single entity $R$. 
So, $\T$ consists of $2 \times 2$ blocks, and the integrand in \refeqn{ECasimir} reduces to $\tr\ln\left(\tI-\T_A\tGM\T_R\tGM\right)$. 
Merging the components of $R$ poses no conceptual difficulty given that the operators are expressed in a position basis, while an actual computation of the force between $A$ and $R$ would remain a daunting task.
If object $A$ is moved infinitesimally by vector $\vecd$, the Laplacian of the energy is given by
\begin{align}
\left.\nabla^2_{\vecd} \,\mathcal{E}\right|_{\vecd=0} & =
-\frac{\hbar c}{2\pi} \int_0^\infty d\ka \,\,\tr
\Big[
2 n_M^2(ic\kappa) \ka^2 \tfrac{\T_A \tGM \T_R \tGM}{\tI-\T_A \tGM \T_R \tGM} \labeleqn{line1} \\
& + 2 \T_A \bnabla \tGM \T_R \left( \bnabla \tGM \right)^T \tfrac{\tI}{\tI-\T_A \tGM \T_R \tGM}  \labeleqn{line2} \\
& + 2 \T_A \bnabla \tGM \T_R \tGM \tfrac{\tI}{\tI-\T_A \tGM \T_R \tGM} \labeleqn{line3} \\
& \cdot
\left(\T_A \bnabla \tGM \T_R \tGM + \T_A \tGM \T_R \left( \bnabla \tGM \right)^T \right) \tfrac{\tI}{\tI-\T_A \tGM \T_R \tGM}  \Big] \nonumber \, .
\end{align}
After displacement of object $A$, the Green's function  multiplied by $\T_A$ on the left and $\T_R$ on the right $(\T_A\tGM\T_R)$ becomes $\tGM(ic\kappa,\vecx+\vecd,\vecx')$, while that multiplied by $\T_R$ on the left and $\T_A$ on the right $(\T_R\tGM\T_A)$ becomes $\tGM(ic\kappa,\vecx,\vecx'+\vecd)$. 
The two are related by transposition, and indicated by $\bnabla \tGM(ic\kappa,\vecx,\vecx') = \bnabla_{\vecd} \tGM(ic\kappa,\vecx+\vecd,\vecx')|_{\vecd=0}$ and $\left(\bnabla \tGM(ic\kappa,\vecx,\vecx')\right)^T = \bnabla_{\vecd} \tGM(ic\kappa,\vecx,\vecx'+\vecd)|_{\vecd=0}$ in the above equation.
In the first line we have substituted $n_M^2(ic\kappa) \kappa^2 \tGM$ for $\nabla^2 \tGM$; the two differ only by derivatives of $\delta$--functions which vanish since $\tGM\left(ic\kappa,\vecx,\vecx'\right)$ is evaluated with $\vecx$ in one object and $\vecx'$ in another. 
In expressions not containing inverses of $\T$-operators, we can extend the domain of all operators to the entire space: 
$\T_J(ic\kappa,\vecx,\vecx')=0$ if $\vecx$ or $\vecx'$ are not on object $J$ and thus operator multiplication is unchanged. 

To determine the signs of the various terms in $\left.\nabla^2_{\vecd} \,\mathcal{E}\right|_{\vecd=0}$, an analysis similar to Ref.~\cite{Kenneth06}{ can be performed. }{Consequently, }the Laplacian of the energy is{ found to be} smaller than or equal to zero as long as {both $\T_A$ and $\T_R$ are either positive or negative semidefinite for all imaginary frequencies \footnote{In practice, $\T_A$ and $\T_R$ suffice to have the same sign over
the frequencies, which contribute most to the integral (or the sum) in
\refeqn{ECasimir}.}}.
{The eigenvalues of $\T_J$, defined in \refeqn{Tem}, on the other hand, are greater or smaller than zero} depending on the sign $s^J$ of $\V_J$, since 
\begin{equation}
 \T_J = s^J \sqrt{s^J \V_J}\frac{\tI}{\tI+ s^J \sqrt{s^J \V_J} \tGM \sqrt{s^J \V_J}} \sqrt{s^J \V_J}\,.
 \end{equation} 

We are left to find the sign of the potential, 
\begin{align}
\mathbb{V}_J(ic\kappa,\vecx)&=\tI\,\ka^2\left(\ep_J(ic\ka,\vecx)-\ep_M(ic\ka)\right)\nonumber\\
&+ \curl\left(\mu_J^{-1}(ic\ka,\vecx)-\mu^{-1}_M(ic\ka)\right)\curl\,,
\end{align}
 of
the object $A$, and the compound object $R$
\footnote{The first curl in the operator $\mathbb{V}_J$ results from an integration by parts. It is understood that it acts on the wave function multiplying $\mathbb{V}_J$ from the left.}.
The sign is determined
by the relative permittivities and permeabilities of the objects and
the medium: If $\ep_J(ic\ka,\vecx) > \ep_M(ic\ka)$ and
$\mu_J(ic\ka,\vecx) \leq \mu_M(ic\ka)$ hold for all $\vecx$ in object
$J$, the potential $\V_J$ is positive.  If the opposite inequalities
are true, $\V_J$ is negative.  The curl operators surrounding the
magnetic permeability do not influence the sign, as in computing an
inner product with $\V_J$ they act symmetrically on both sides.  For
vacuum $\ep_M=\mu_M=1$, and material response functions
$\ep(ic\ka,\vecx)$ and $\mu(ic\ka,\vecx)$ are analytical continuations
of the permittivity and permeability for real
frequencies~\cite{LandauLifshitz8}.  While
$\ep(ic\ka,\vecx)> 1$ for positive $\ka$, there are no
restrictions other than positivity on $\mu(ic\ka,\vecx)$.
(For non-local and non-isotropic response, various inequalities must be generalized to the tensorial operators $\overleftrightarrow{\boldsymbol{\epsilon}}(ic\kappa,\vecx,\vecx')$ and $\overleftrightarrow{\boldsymbol{\mu}}(ic\kappa,\vecx,\vecx')$.)

Thus, levitation is not possible for collections of objects characterized by $\ep_J(ic\kappa,\vecx)$ and $\mu_J(ic\kappa,\vecx)$ falling
into one of the two classes described earlier, i) $\ep_J/\ep_M>1$ and $\mu_J/\mu_M\leq 1$ (positive
$\V_J$ and $\T_J$), or ii) $\ep_J/\ep_M<1$ and $\mu_J/\mu_M\geq 1$ with
(negative $\V_J$ and $\T_J$).
(Under these conditions parallel slabs attract.) The frequency and space dependence of the functions has been suppressed in these inequalities. In vacuum,
$\ep_M(ic\kappa)=\mu_M(ic\kappa)=1$; since $\ep(ic\kappa,\vecx)> 1$ and
the magnetic response of ordinary materials is typically
negligible~\cite{LandauLifshitz8}, one concludes that
stable equilibria of the Casimir force do not exist. If objects $A$
and $R$, however, belong to different categories --- under which
conditions the parallel plate force is repulsive --- then the terms
under the trace in lines \refeq{line1} and \refeq{line2} are negative.
The positive term in line \refeq{line3} is typically smaller than the
first two, as it involves higher powers of $\T$ and $\tGM$.  In this
case stable equilibrium is possible, as demonstrated recently for a
small inclusion within a dielectric filled cavity \cite{Rahi09-2}.
For the remaining two combinations of inequalities involving $\ep_J/\ep_M$ and $\mu_J/\mu_M$ the sign of $\V_J$ cannot be determined a priori. 
But for realistic distances between objects and the corresponding frequency ranges, the magnetic susceptibility is negligible for ordinary materials, and the inequalities involving $\mu$ can be ignored.

In summary, the instability theorem applies to all cases where the
coupling of the EM field to matter can be
described by response functions $\epsilon$ and $\mu$, which may vary
continuously with position and frequency. 
Obviously, for materials which at a microscopic level cannot be described by such response functions, e.g., because of magneto-electric coupling, our theorem is not applicable.

Even complicated arrangements of materials obeying the above conditions are subject to the instability constraint. For example, metamaterials,
incorporating arrays of micro-engineered circuity mimic, at certain
frequencies, a strong magnetic response, and have been discussed as
candidates for Casimir repulsion across vacuum.
(References~\cite{Rosa08,Rosa09} critique repulsion from dielectric/metallic based metamaterials,
in line with our following arguments.)
In our treatment, in accord with the usual electrodynamics of macroscopic media, the materials are characterized by $\ep(ic\kappa,\vecx)$ and $\mu(ic\kappa,\vecx)$ at mesoscopic scales.
In particular, chirality and large magnetic response in metamaterials are achieved by patterns made from ordinary metals and dielectrics with well-behaved $\ep(ic\kappa,\vecx)$ and $\mu(ic\kappa,\vecx)\approx 1$ at \emph{short} scales. The interesting EM responses 
merely  appear when viewed as `effective' or `coarse grained'.

Clearly, the coarse-grained response functions, which are conventionally employed to describe metamaterials, should produce, in their region of validity, the same scattering amplitudes as the detailed mesoscopic description.
Consequently, as long as the metamaterial can be described by $\ep(ic\kappa,\vecx)$ and $\mu(ic\kappa,\vecx)\approx 1$, the eigenvalues of the $\T$ operators are constrained as described above, and hence subject to the instability theorem.
Thus, the proposed use of chiral metamaterials in reference~\cite{Zhao09} cannot lead to stable equilibrium since the structures are composites of metals and dielectrics.
Finally, we note that instability also excludes repulsion between two objects that obey the above conditions, if one of them is an infinite flat plate with continuous translational symmetry: Repulsion would require that the energy as a function of separation from the slab should have $\partial^2_d \mathcal{E}>0$ at some point since the force has to vanish at infinite separation.
A metamaterial does not have continuous translational symmetry at short length scales but this symmetry is approximately valid in the limit of large separations (long wavelengths), where the material can be effectively described as a homogeneous medium.
At short separations lateral displacements might lead to repulsion that, however, must be compatible with the absence of stable equilibirum.